\documentclass[12pt,preprint]{aastex}
\usepackage{amsmath}
\usepackage{txfonts}
\usepackage{color}
\usepackage{fancybox}

\newcommand{\mr}[1]{\ensuremath{\mathrm{#1}}}
\newcommand{\ux}[2]{\ensuremath{\mr{^{#1} #2}}}
\newcommand{\pen}{\mr{p(e^{-},\nu_{e})n}}
\newcommand{\nep}{\mr{n(e^{+},\overline{\nu_{e}})p}}

\begin{document}

\title{Freeze-out yields of radioactivities in core-collapse supernovae}

\author{
Georgios Magkotsios\altaffilmark{1,2,3},
F. X. Timmes\altaffilmark{2,3},
Michael Wiescher\altaffilmark{1,3}
       }
\altaffiltext{1}{Department of Physics, University of Notre Dame,
                 Notre Dame, IN 46556, USA}
\altaffiltext{2}{School of Earth and Space Exploration,
                 Arizona State University,
                 Tempe, AZ 85287, USA}
\altaffiltext{3}{The Joint Institute for Nuclear Astrophysics,
                 Notre Dame, IN 46556, USA}

\begin{abstract}

We explore the nucleosynthesis trends from two mechanisms during
freeze-out expansions in core-collapse supernovae. The first
mechanism is related to the convection and instabilities within
homogeneous stellar progenitor matter that is accreted through the
supernova shock. The second mechanism is related to the impact of
the supersonic wind termination shock (reverse shock) within the
tumultuous inner regions of the ejecta above the proto-neutron star.
Our results suggest that isotopes in the mass range $12\leqslant
A\leqslant122$ which are produced during the freeze-out expansions
may be classified in two families. The isotopes of the first family
manifest a common mass fraction evolutionary profile, whose specific
shape per isotope depends on the characteristic transition between
two equilibrium states (equilibrium state transition) during each
type of freeze-out expansion. The first family includes the majority
of isotopes in this mass range. The second family is limited to
magic nuclei and isotopes in their locality which do not sustain any
transition, become nuclear flow hubs, and dominate the final
composition. We use exponential and power-law adiabatic profiles to
identify dynamic large-scale and small-scale equilibrium patterns
among nuclear reactions. A reaction rate sensitivity study
identifies those reactions which are crucial to the synthesis of
radioactivities in the mass range of interest. In addition, we
introduce non-monotonic parameterized profiles to probe the impact
of the reverse shock and multi-dimensional explosion asymmetries on
nucleosynthesis. Cases are shown where the non-monotonic profiles
favor the production of radioactivities. Non-monotonic freeze-out
profiles involve longer non-equilibrium nucleosynthesis intervals
compared to the exponential and power-law profiles, resulting in
mass fraction trends and yield distributions which may not be
achieved by the monotonic freeze-out profiles.

\end{abstract}


\section{INTRODUCTION}
\label{sec:intro}

Despite the progress in core-collapse supernova theory during the
past 20 years, the details of a fully self-consistent explosion and
the nucleosynthesis are not yet fully understood. Three-dimensional
supernova simulations with energy-dependent neutrino transport are
necessary to clarify the growth of convective and Rayleigh--Taylor
instabilities, the possibility of non-radial, non-axisymmetric
instability modes, and the development of local fluid vortices
\citep{janka_2007_aa}. Further mixing processes near the
proto-neutron star surface that could enhance the neutrino heating
behind the shock include doubly diffusive instabilities
\citep{bruenn_2004_aa}, neutrino-bubble instabilities
\citep{socrates_2005_aa}, and magnetic buoyancy instabilities
\citep{wilson_2005_aa}. The neutrino heating results in the outflow
of baryonic matter (wind) from the surface of the proto-neutron star
\citep{duncan_1986_aa}. This neutrino-driven wind interacts with the
more slowly moving, earlier supernova ejecta forming a wind
termination shock, which is also termed as reverse shock
\citep{burrows_1995_aa,janka_1995_aa}. The position of the reverse
shock depends on the wind velocity and the thermodynamic conditions
of the supernova ejecta which the wind collides with
\citep{arcones_2007_aa}, while its boundary is strongly deformed to
a non-spherical shape in multi-dimensional simulations
\citep{arcones_2011_aa}.

The interaction of the supernova shock with the stellar progenitor
layers and the reverse shock with the mass layers above the
proto-neutron star results in sudden entropy increases which trigger
nucleosynthesis in the ashes of pre-collapse nuclear burning. The
expanding material then cools down until nuclear reactions
freeze-out. The temperature and density trajectories of the
expanding ejecta located initially outside the range of convective
and mixing processes are monotonically decreasing. However, certain
thermodynamic trajectories may be non-monotonic and can affect the
abundances' evolution in non-trivial ways. Possible factors to
result in non-monotonic trajectories include the explosion
energetics \citep{nakamura_2001_aa}, explosion asymmetries in
multi-dimensional simulations
\citep{nagataki_1997_aa,nagataki_1998_aa}, rotating progenitors
\citep{fryer_2000_aa}, double explosion scenaria
\citep{fryer_2006_ab}, and the effect of the reverse shock
\citep{arcones_2007_aa,arcones_2011_aa}. The yields from
non-monotonic trajectories are usually different compared to
compositions from monotonic trajectories.

Certain studies have explored the impact of the supernova shock and
the reverse shock on nucleosynthesis processes such as the
$\alpha$-rich and neutron-rich freeze-out
\citep{woosley_1992_aa,witti_1994_aa}, the $\nu$$p$-process
\citep{frohlich_2006_aa,pruet_2006_ab}, the $rp$-process
\citep{wanajo_2006_aa}, and the $r$-process
\citep{takahashi_1994_aa}. For the $r$-process specifically, the
impact of the reverse shock has been studied under various wind
schemes which include subsonic ``breeze'' solutions with constant
pressure boundaries \citep{terasawa_2002_aa}, supersonic winds with
fixed asymptotic temperature \citep{wanajo_2002_aa}, and two-phase
outflow models \citep{kuroda_2008_aa}, including reverse shock
temperature ranges which could trigger the ``cold'' $r$-process
\citep{wanajo_2007_aa} or the classical $r$-process
\citep{kratz_2007_aa}. However, integrated nucleosynthesis based on
those simulations shows that no heavy $r$-process elements can be
produced \citep{arcones_2011_ab}. In addition, reaction rate
sensitivity studies have been performed to constrain uncertainties
related to the $p$-process \citep{rapp_2006_aa}, $\nu$$p$-process
\citep{wanajo_2011_aa}, freeze-out expansions from nuclear
statistical equilibrium (henceforth NSE) or quasi-static equilibrium
(henceforth QSE) in
\citet{the_1998_aa,hoffman_2010_aa,magkotsios_2010_aa}, and
hydrostatic burning \citep{tur_2010_aa}. Sensitivity studies add
detail to the understanding of the microscopic mechanisms which
drive the related processes, because these mechanisms may not be
easily identified for nominal values of the reaction rates or other
related parameters.

Yields of radioactivities as inferred from observations and presolar
signatures (calcium and aluminum-rich inclusions, inter-planetary
dust particles, and stellar grains) add complexity and raise new
challenges for the explosion mechanism. Isotopic anomalies in
presolar signatures indicate discrepancies between measured or
inferred isotopic ratios and current supernova theory. The gamma-ray
diffusive emission pattern of \ux{26}{Al} indicates that massive
stars are its most probable source
\citep{leising_2009_aa,limongi_2006_aa,tur_2010_aa}, given its
correlation to \ux{60}{Fe} \citep{timmes_1995_aa,limongi_2006_ab}.
Yet, the synthesis of \ux{60}{Fe} is largely dependent on the
reaction rates producing it \citep{leising_2009_aa}. \ux{41}{Ca}
found in SiC grains is likely to have common origin with \ux{26}{Al}
\citep{sahijpal_1998_aa}, with a small fraction of \ux{41}{Ca}
possibly originating from supernova explosions
\citep{nittler_2008_aa}. It is argued that \ux{53}{Mn} was uniformly
distributed in the early solar nebula \citep{yamashita_2010_aa}.
Although \ux{53}{Mn} is produced in significant amount in
core-collapse supernovae, it does not emit $\gamma$-rays which would
be easily detectable, but only X-rays which require significant
accumulation in the interstellar medium for successful detection
\citep{leising_2009_aa}. \ux{79}{Se} also lacks a decay scheme by
$\gamma$-rays, although its detection in presolar grains is
possible. Heavier isotopes such as \ux{93}{Zr}
\citep{lugaro_2003_aa}, \ux{92}{Nb}, and \ux{97}{Tc} are believed to
be produced primarily in asymptotic giant branch (AGB) stars,
although we demonstrate cases where they may be produced in
significant amounts during supernova explosions.

In this study, we use monotonic adiabatic profiles to quantify
nucleosynthesis trends caused by equilibrium patterns among nuclear
reactions. In addition, we introduce non-monotonic parameterized
expansion profiles to simulate the impact of two mechanisms which
may result in non-monotonic evolution for the local temperature and
density evolution. The first mechanism involves multi-dimensional
effects of the explosion following the passage of the supernova
shock through the homogeneous layers of progenitor gas, and the
second mechanism involves the impact of the abrupt deceleration by
the reverse shock to the supersonic wind originating from the
surface of the proto-neutron star. The mechanisms are independent of
each other, because a contact discontinuity prevents mixing between
the accumulated wind matter and the dense layer of shock-accelerated
progenitor gas \citep{arcones_2007_aa}. The non-monotonic profiles
result in yield profiles which may not be achieved by the monotonic
expansions. These results indicate the importance of
multi-dimensional supernova simulations on nucleosynthesis within
the tumultuous stellar mass layers. We demonstrate cases where the
non-monotonic profiles favor the production of radioactivities, and
conclude that under certain circumstances non-monotonic profiles may
increase the likelihood of detecting radioactivities in observed
supernovae.

Section \ref{sec:old_work} is a summary of our previous work
\citep{magkotsios_2010_aa}. In Section \ref{sec:profiles} we present
the formalism of the parameterized thermodynamic trajectories.
Section \ref{sec:monotonic} considers general trends of isotopes in
the mass range $12\leqslant A\leqslant122$ from our exponential and
power-law trajectories. Sensitivities to reaction rate values
related to the synthesis of radioactivities are discussed in Section
\ref{sec:reactions}. The impact of non-monotonic profiles on
nucleosynthesis is discussed in Section \ref{sec:non-monotonic}. We
conclude our discussion with a summary of our new results in Section
\ref{sec:summary}.

Our nomenclature is based on the following conventions. The
temperature of the material is expressed by the pure number $T_9 =
T/(10^9 \ {\rm K})$. The electron fraction, or the total proton to
nucleon ratio is $Y_{e} = \sum_{i}Z_{i}Y_{i} =
\sum_{i}{Z_{i}}/{A_{i}}\ X_{i}$. We define ``nuclear flow'' to mean
the instantaneous rate of change of isotope $i$'s molar abundance
with time ($dY_{i}/dt$) due to a given nuclear reaction
\citep{iliadis_2007_aa}. For any single reaction linking isotope $i$
with isotope $j$ there is a forward flow, a reverse flow, and a
relative net flow $\phi_{i}$=(forward $-$
reverse)/max(forward,reverse) that measures the equilibrium state of
the reaction.

\section{SYNOPSIS OF PREVIOUS WORK}
\label{sec:old_work}

The current work is a continuation of \citet{magkotsios_2010_aa}. We
provide a synopsis of key notions and results presented in our
previous work which are relevant to our new results.
\citet{magkotsios_2010_aa} used two parameterized expansion profiles
within a parameter space of peak temperatures, peak densities, and
initial electron fraction values, to discuss nucleosynthesis trends
related to \ux{44}{Ti} and \ux{56}{Ni} during freeze-out expansions
from core-collapse supernovae. The initial composition was dominated
by \ux{28}{Si}, with neutrons or protons added to configure the
value of the initial electron fraction $Y_e$. Setting the initial
composition in this manner was not restrictive for the largest part
of our parameter space. The peak conditions established a
large-scale structure composed by nuclear reactions in equilibrium
and the mass fractions within this structure were rearranged to
their equilibrium values early during the evolution. Large-scale
equilibrium structures involve either nuclear statistical
equilibrium (NSE), or global quasi-static equilibrium (global QSE).
During NSE every reaction within a network is in equilibrium, while
a small subset of them may be off equilibrium during QSE. For
decreasing temperature and density values during the expansion the
large-scale QSE cluster dissolves in multiple small-scale QSE
clusters, where each cluster encompasses only a small number of
nuclei with similar masses (local QSE). Further decrease in the
temperature and density results in the cease of nuclear reactions
(freeze-out stage).

Multiple types of freeze-out expansions were identified and
discussed within the parameter space used, such as the normal
freeze-out
\citep{woosley_1973_aa,meyer_1994_aa,meyer_1998_aa,hix_1999_aa}, the
$\alpha$-rich freeze-out \citep{woosley_1973_aa}, the $\alpha$-rich
and proton-rich freeze-out ($\alpha$$p$-rich), the incomplete
silicon burning regime,, the photodisintegration regime the
\mr{(p,\gamma)}-leakage regime for $Y_e>0.5$, the $\alpha$-rich and
neutron-rich freeze-out for $Y_e<0.5$ ($\alpha$$n$-rich), and the
depletion region of yields which separates the incomplete silicon
burning and normal freeze-out regions from the $\alpha$-rich
freeze-out (chasm).

Each type of freeze-out is related to distinct transitions between
two very different equilibrium states. In this work we refer them as
``equilibrium state transitions'' and we abbreviate them as EST.
These transitions between equilibrium states involve multiple
reactions within the QSE cluster which break equilibrium, resulting
in changes to the QSE cluster's size and shape. The scale of the
transitions during freeze-out expansions may be global, such as the
division of a large QSE cluster encompassing almost every isotope in
the network into two large QSE clusters localized within the silicon
and iron groups respectively. Alternatively, the scale of the
transitions may be local, involving only a small set of nuclear
reactions which form small-scale clusters such as equilibrium chains
of $\mr{(p,\gamma)}$ or $\mr{(n,\gamma)}$ reactions interacting with
$\alpha$-captures or $\mr{(p,n)}$ reactions, respectively. ESTs are
entropy driven, where the temperature sets an approximate threshold
for a transition, while the density at the threshold temperature
determines whether the transition takes place or not. Electron
fraction variations, the expansion timescale, and key reaction rates
control the local equilibrium patterns which shape the locus of each
region.

\section{PARAMETERIZED PROFILES}
\label{sec:profiles}

The nucleosynthesis calculations implement mature reaction network
solvers \citep{timmes_1999_ab,fryxell_2000_aa} and utilize the 489
and 3304 isotope networks described in \citet{magkotsios_2010_aa}.
The 489 isotope network has been expanded to 553 isotopes to include
a sufficient amount of isotopes near the magic number 50, and is
listed in Table \ref{tab:net553}. The 553 isotope network includes
two separate isomers for \ux{26}{Al}, \ux{26}{Al}\mr{_g} for the
ground state and \ux{26}{Al}\mr{_m} for the first excited metastable
state. We use three parameterized profiles to model the freeze-out
expansions. All profiles assume that a passing supernova shock wave
heats material to a peak temperature $T_0$ and compresses the
material to a peak density $\rho_0$. This material then expands and
cools down (freezes out) until the temperature and density are
reduced to the extent that nuclear reactions cease.

The first two parameterized profiles involve monotonic temperature
and density decrease, following a constant $T^3/\rho$ evolution
which implies constant radiation entropy in suitable limits. The
first profile is the exponential expansion
\citep{hoyle_1964_aa,fowler_1964_aa}
\begin{equation}
T(t)=T_{0} \exp(-t/3 \tau) \quad \rho(t) = \rho_{0} \exp(-t/\tau)
\label{eq:Tad_rhoad}
\end{equation}
with a static free-fall timescale for the expanding ejecta
\begin{equation}
\tau=(24\pi G\rho_{0})^{-1/2} \approx 446/\rho_{0}^{1/2} \ {\rm s.}
\label{eq:AD_timescale}
\end{equation}
The exponential profile has been used extensively in the past to
explore yield trends and their sensitivity to reaction rates or
electron fraction values
\citep{woosley_1973_aa,woosley_1992_aa,the_1998_aa,hoffman_2010_aa,magkotsios_2010_aa}.
The second profile is a power-law based on homologous expansion
introduced by \citet{magkotsios_2008_aa,magkotsios_2010_aa}
\begin{equation}
T(t)=\frac{T_{0}}{ct+1} \quad \rho(t)=\frac{\rho_{0}}{(ct+1)^{3}}
\enskip , \label{eq:Tpl_rhopl}
\end{equation}
where the coefficient $c=2$ s$^{-1}$ is chosen to mimic trajectories
taken from core-collapse simulations.

\begin{figure}[htp]
\includegraphics[width=0.95\textwidth]{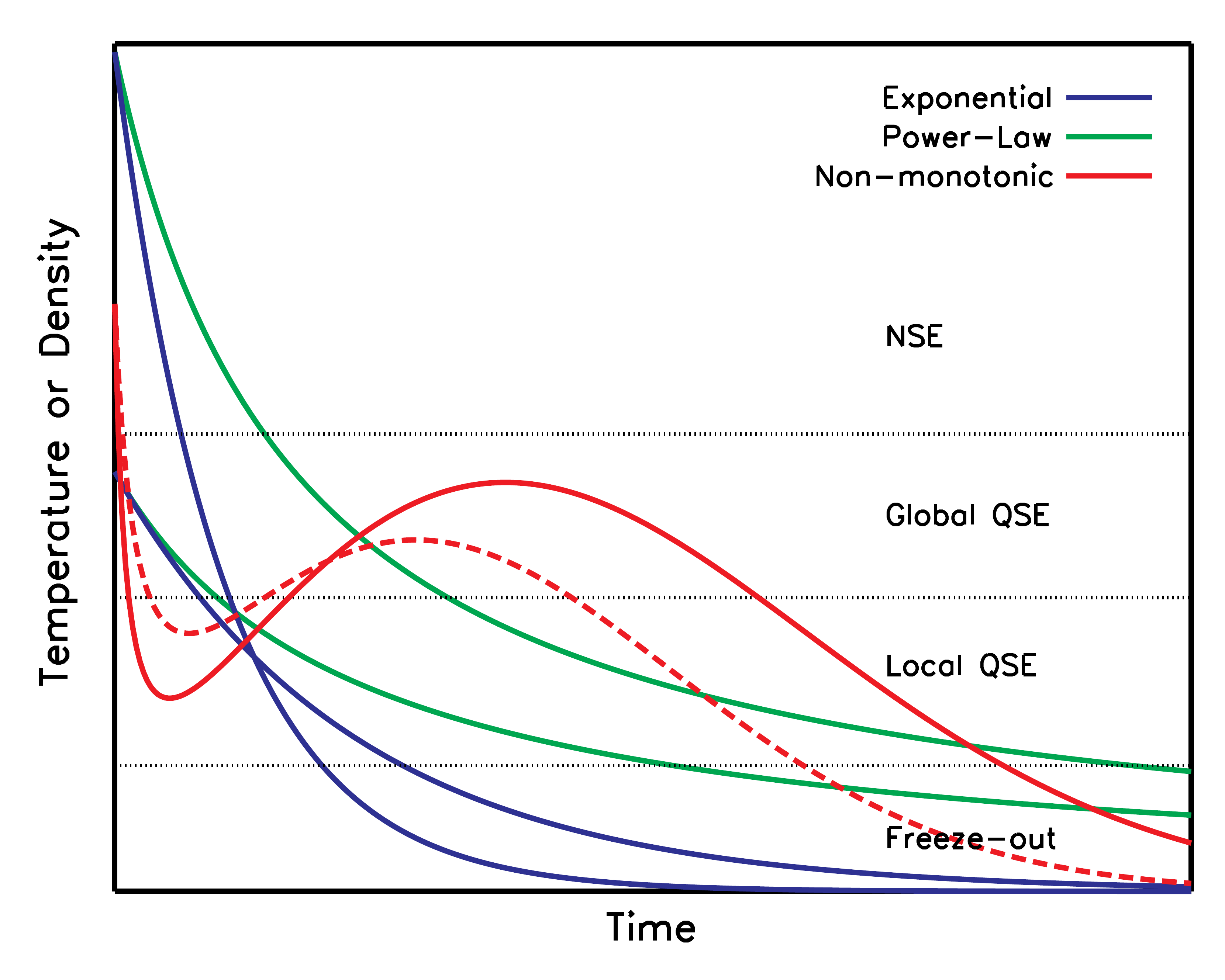}
\caption{Schematic temperature or density evolution for the
exponential, power-law and non-monotonic profiles. Passages through
different burning regimes for various peak conditions are indicated.
The dashed red curve illustrates the impact of variations to the
values of the local extremum points for the same peak conditions.}
\label{fig:regimes_cartoon}
\end{figure}

Figure \ref{fig:regimes_cartoon} compares the general properties of
the exponential and power-law profiles. For a given initial
condition, the power-law evolution is always slower than the
exponential one. Moreover, the power-law evolution becomes slower
for increasing peak temperature and density values. The differences
in these two profiles affect the final yields as material traverses
different burning regimes on different timescales. The figure also
depicts the NSE, global QSE, local QSE, and final freeze-out burning
regimes. The exponential and power-law trajectories are chosen so
that they generally bound the temperature and density trajectories
of hydrodynamic particles from spherically symmetric and
two-dimensional explosion models.

We introduce non-monotonic profiles to simulate possible
thermodynamic conditions in the ejecta from different mass layers
within the star. Non-monotonic temperature and density evolution may
arise due to convection and instabilities following the heating of
homogeneous progenitor mass by the supernova shock, or by the
termination shock to the supersonic wind within the inner mass
layers above the proto-neutron star. Our profiles involve three
stages to model such possible effects (Figure
\ref{fig:regimes_cartoon}). During the first stage, the supernova
shock rises the temperature and density values of the material
traversed, and the material is allowed to expand while it cools and
rarefies. During the second stage we introduce a simplified approach
to model the effect of multi-dimensional asymmetries to the
explosion or the reverse shock within a parameterized expansion
profile. This stage involves a contraction phase which rises the
temperature and density linearly in time to a local maximum. The
third stage involves an exponential freeze-out, because the material
is assumed to be part of the ejecta and eventually escape from the
star. Contrary to the exponential and power-law cases, no explicit
assumption is made about holding $T^3/\rho$ constant.

Each temperature and density trajectory has a peak value followed by
a local minimum and then a local maximum. The local minimum and
maximum values for the case of the reverse shock depend on the wind
velocity and the deformed boundary of the shock near the supernova
ejecta which the wind collides with
\citep{arcones_2007_aa,arcones_2011_aa}. For the case of our
parameterized expansion profiles, we choose the local extremum
values for temperature and density randomly from a uniform
distribution. Each value of a local extremum point is considered to
be independent and identically distributed. This is a reasonable
approach, because we aim to study the key nucleosynthesis trends
within the tumultuous inner layers of the ejecta, and we attempt to
simplify the hydrodynamic evolution for this purpose.

The differential equations for the non-monotonic temperature and
density trajectories are
\begin{align}
\frac{dT}{dt} &= -T_{0}\exp(-k_{_T} t)\biggl[k_{_T} b_{_T} t +
\frac{k_{_T}}{1+a_{_T} t} + \frac{a_{_T}}{(1+a_{_T} t)^2} -
b_{_T}\biggr]
\label{eq:temp_asym_ode}\\
\frac{d\rho}{dt} &= -\rho_{0}\exp(-k_\rho t)\biggl[k_\rho b_\rho t +
\frac{k_\rho}{1+a_\rho t} + \frac{a_\rho}{(1+a_\rho t)^2} -
b_\rho\biggr] \label{eq:rho_asym_ode}
\end{align}
and their solutions are
\begin{align}
T(t) &= T_{0}\exp(-k_{_T} t)\biggl[\frac{1}{1+a_{_T} t} + b_{_T}
t\biggr] \label{eq:temp_asym}\\
\rho(t) &= \rho_{0} \exp(-k_\rho t)\biggl[\frac{1}{1+a_\rho t} +
b_\rho t\biggr] \enskip , \label{eq:rho_asym}
\end{align}
where the parameters $a_i$, $b_i$ and $k_i$ may be chosen to control
the local extremum points. For instance, the density minimum is
given approximately by
$\rho_{min}=\rho_0(2\sqrt{a_{\rho}b_{\rho}}-b_{\rho})/a_{\rho}$ at
time
$t_{min}=(\sqrt{a_{\rho}}-\sqrt{b_{\rho}})/(a_{\rho}\sqrt{b_{\rho}})$,
while the maximum is given by $\rho_{max}=\rho_0/(e\;k_{\rho})$ at
time $t_{max}=1/k_{\rho}$. The red curves of Figure
\ref{fig:regimes_cartoon} show the profile's general trends compared
to the exponential and power-law expansions. The ascending
trajectory is focused on QSE and local equilibrium stages. We ensure
that the local maximum for the temperature does not exceed the NSE
threshold, otherwise the preceding part of the trajectory would not
impact the evolution. The local minimum for the temperature ranges
from the freeze-out temperature $T_9 = 0.01$ until the peak
temperature. The subsequent local maximum for the temperature ranges
from the value of the local minimum until the value $T_9=4$. Thus, a
non-monotonic evolution is guaranteed without the re-establishment
of NSE. The range of the extremum point values for the density
profile is less constrained and even monotonic profiles may arise.
The minimum spans the range $10^5$ g cm$^{-3}$ $<\rho_{min}<\rho_0$
and the range for the maximum is $10^6$ g cm$^{-3}$
$<\rho_{max}<10^{9}$ g cm$^{-3}$.

\section{YIELDS FROM THE EXPONENTIAL AND POWER-LAW PROFILES}
\label{sec:monotonic}

The exponential and the power-law expansion profiles used in this
work set the basis for quantifying the details of nucleosynthesis
mechanisms. Their low computational cost allows the monitoring of
large-scale and small-scale equilibrium patterns among nuclear
reactions in time, which is a powerful tool for identifying
microscopic components that affect the composition of the ejecta.

Monotonic profiles are probes for multiple burning regimes. For
instance, during subsonic outflows (neutrino-driven ``breezes'')
from the proto-neutron star, the flow merges smoothly with the
denser shell of ejecta behind the outgoing supernova shock resulting
in monotonic evolution for both the temperature and density of the
ejecta \citep{otsuki_2000_aa,terasawa_2002_aa,arcones_2007_aa}. For
cases where the temperature increase imparted by the reverse shock
during supersonic winds is above the NSE threshold $T_9\sim5$, the
nuclear abundances acquire NSE values and the previous thermodynamic
history of the ejecta is irrelevant to the nucleosynthesis evolution
following the temperature jump. Temperature increases above the NSE
threshold may also occur when the homogeneous progenitor matter is
traversed by the supernova shock. Monotonic profiles are suitable
probes for cases where mixing processes below the NSE threshold are
either absent or negligible.

\begin{figure}[htp]
\includegraphics[width=0.95\textwidth]{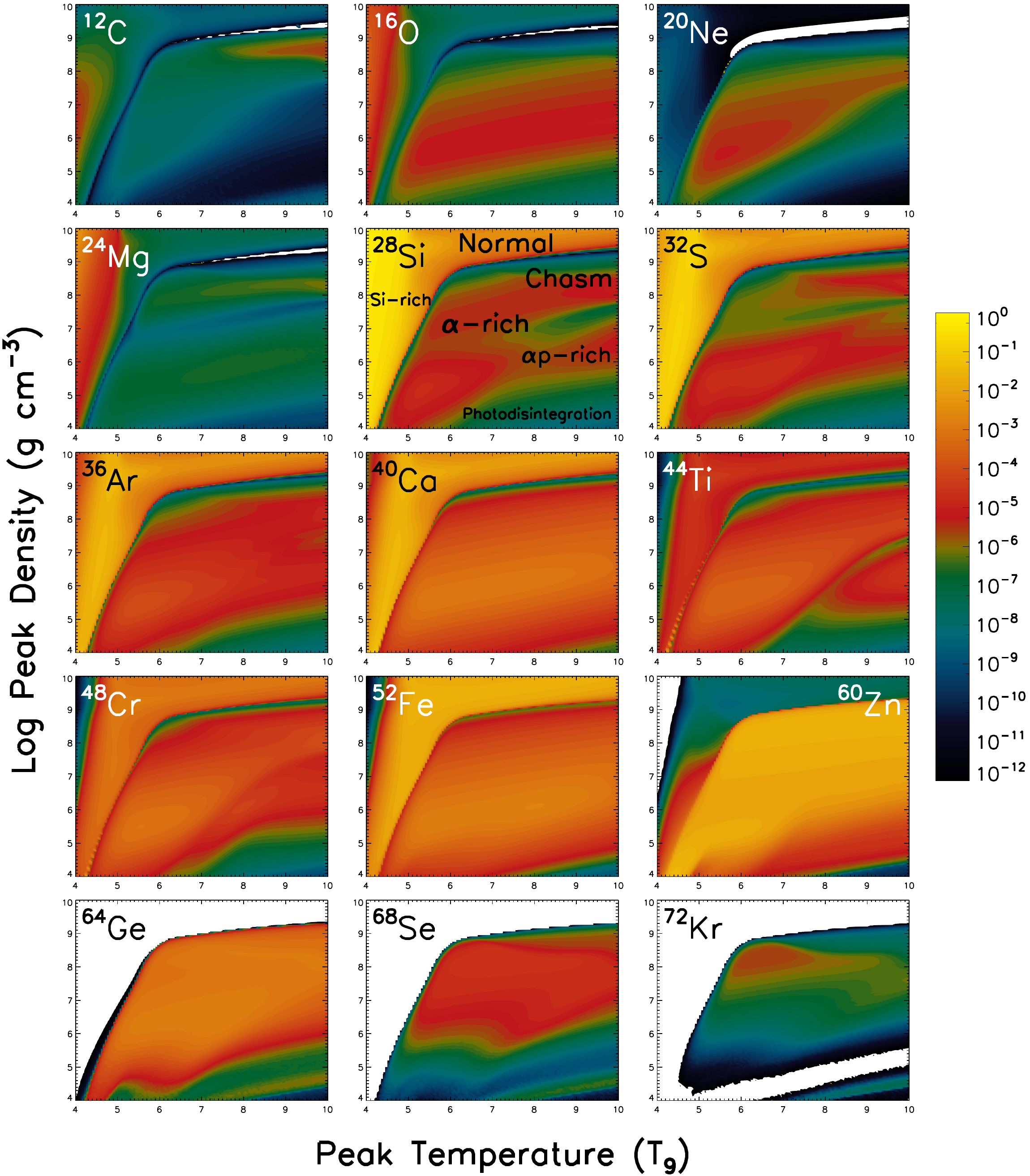}
\caption{Final mass fractions of the $\alpha$-chain isotopes
(\ux{56}{Ni} omitted) in the peak temperature--density plane for the
exponential profile at $Y_{e} = 0.5$. The white colored space
corresponds to values below the color scale shown. From left to
right, the first row corresponds to \ux{12}{C}, \ux{16}{O} and
\ux{20}{Ne}, the second row corresponds to \ux{24}{Mg}, \ux{28}{Si}
and \ux{32}{S}, the third row corresponds to \ux{36}{Ar},
\ux{40}{Ca} and \ux{44}{Ti}, the fourth row corresponds to
\ux{48}{Cr}, \ux{52}{Fe} and \ux{60}{Zn}, and the fifth row
corresponds to \ux{64}{Ge}, \ux{68}{Se} and \ux{72}{Kr}.}
\label{fig:contour_AD1_ye0500_a-chain}
\end{figure}

Figure \ref{fig:contour_AD1_ye0500_a-chain} shows the final mass
fractions of isotopes along the $\alpha$-chain for initially
symmetric matter ($Y_e=0.5$). The temperature--density planes
include the full range of peak conditions within our parameter
space. With the exception of \ux{56}{Ni} (not shown), the
topological structure of all planes is similar, and is marked by
distinct regions. These regions are labeled in the \ux{28}{Si}
contour plot for the symmetric case, each region corresponding to a
type of freeze-out expansion. Further types of freeze-out expansions
are manifested for initial electron fraction $Y_e\neq0.5$ (see
analysis below). The aggregate range of isotopes produced by all
identified freeze-out types is in the mass range $12\leqslant
A\leqslant122$, including in addition the free neutrons, protons,
and $\alpha$-particles. The structure of the temperature--density
plane is very similar among the majority of the isotopes in this
mass range. We classify into a family the isotopes within the mass
range $12\leqslant A\leqslant122$ whose temperature--density plane
features a region-divided structure (henceforth the ``first family''
of isotopes) and explore the common features that these isotopes
share. The $\alpha$-chain isotopes shown in Figure
\ref{fig:contour_AD1_ye0500_a-chain} belong to the first family.

The mass fraction profiles per region for the isotopes in Figure
\ref{fig:contour_AD1_ye0500_a-chain} and the first family overall
are similar, indicating that within a region the isotopes of this
family are produced by the same mechanism. The mass fraction profile
similarities arise from the initial formation of a large-scale QSE
cluster during the freeze-out processes. Within the cluster all
nuclei are interconnected with reactions in equilibrium, and mass
fraction values are determined by the temperature, density and
electron fraction variations based on minimization principles of the
Helmholtz free energy \citep{seitenzahl_2008_aa}. Subsequent ESTs
alter the shape of the QSE cluster and eventually the cluster
dissolves. The precise locus of the regions in the
temperature--density plane for an isotope of the first family
depends on the local equilibrium patterns near the isotope while the
large QSE cluster dissolves.

\begin{figure}[htp]
\includegraphics[height=0.8\textheight]{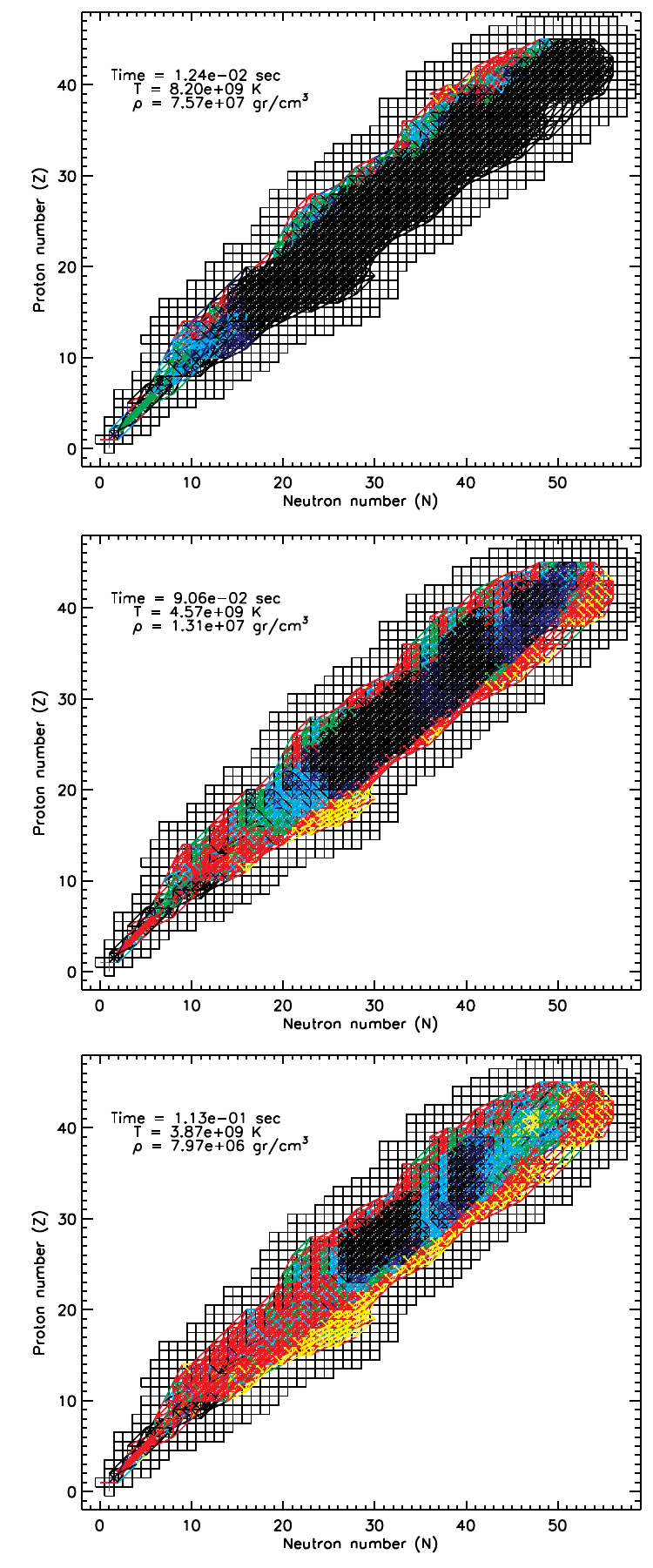}
\caption{QSE cluster motion in the chart of nuclides for an
$\alpha$-rich freeze-out. The QSE cluster remnant condenses near the
magic number 28. Each colored line corresponds to a nuclear reaction
and indicates the level of nuclear flow transferred between the
isotopes connected. Normalized flows $\phi$ are colored black for $0
\le \phi < 0.01$, navy for $0.01 \le \phi < 0.05$, blue for $0.05
\le \phi < 0.1$, cyan for $0.1 \le \phi < 0.4$, green for $0.4 \le
\phi < 0.8$, red for $0.8 \le \phi < 1.0$, and yellow for
$\phi$=1.0. Small $\phi$ values indicate reactions in equilibrium,
while $\phi$=1.0 implies pure one-way nuclear flow transfer.  }
\label{fig:charts_a-rich}
\end{figure}

Figure \ref{fig:charts_a-rich} illustrates the upward shifting in
mass of the QSE cluster \citep{meyer_1998_aa}. The QSE cluster
remnant condenses around the magic number 28, and nuclei in this
small group tend to dominate the final composition. The isotopes of
the first family which are gradually left outside the QSE cluster
form chains of \mr{(p,\gamma)} reactions in equilibrium along the
isotone lines. The first EST related to these isotopes' exit from
the QSE cluster is signaled at the microscopic level by the
equilibrium break of the $\alpha$-capture reactions linking the
\mr{(p,\gamma)} equilibrium chains. During the $\alpha$$p$-rich
freeze-out, isotopes of the first family sustain a second EST when
certain \mr{(p,\gamma)} reactions in the isotone chain break
equilibrium. These small-scale equilibrium patterns are responsible
for producing eventually the isotopes of the first family from
\ux{12}{C} to the iron peak. On the contrary, the formation of the
chasm for each isotope of the first family results from the
dissolution of the large-scale QSE cluster to two smaller ones. The
first cluster encompasses the silicon group elements and the second
cluster encompasses the iron group elements. The cluster breakage
results in massive flow transfer from the silicon and most of the
iron group isotopes toward a small group of nuclides near the magic
number 28. The flow transfer proceeds until all mass fractions are
depleted, excluding the mass fractions of nuclei around the magic
number 28. These nuclei are produced in large amounts and dominate
the final composition.

\begin{figure}[htp]
\includegraphics[width=0.95\textwidth]{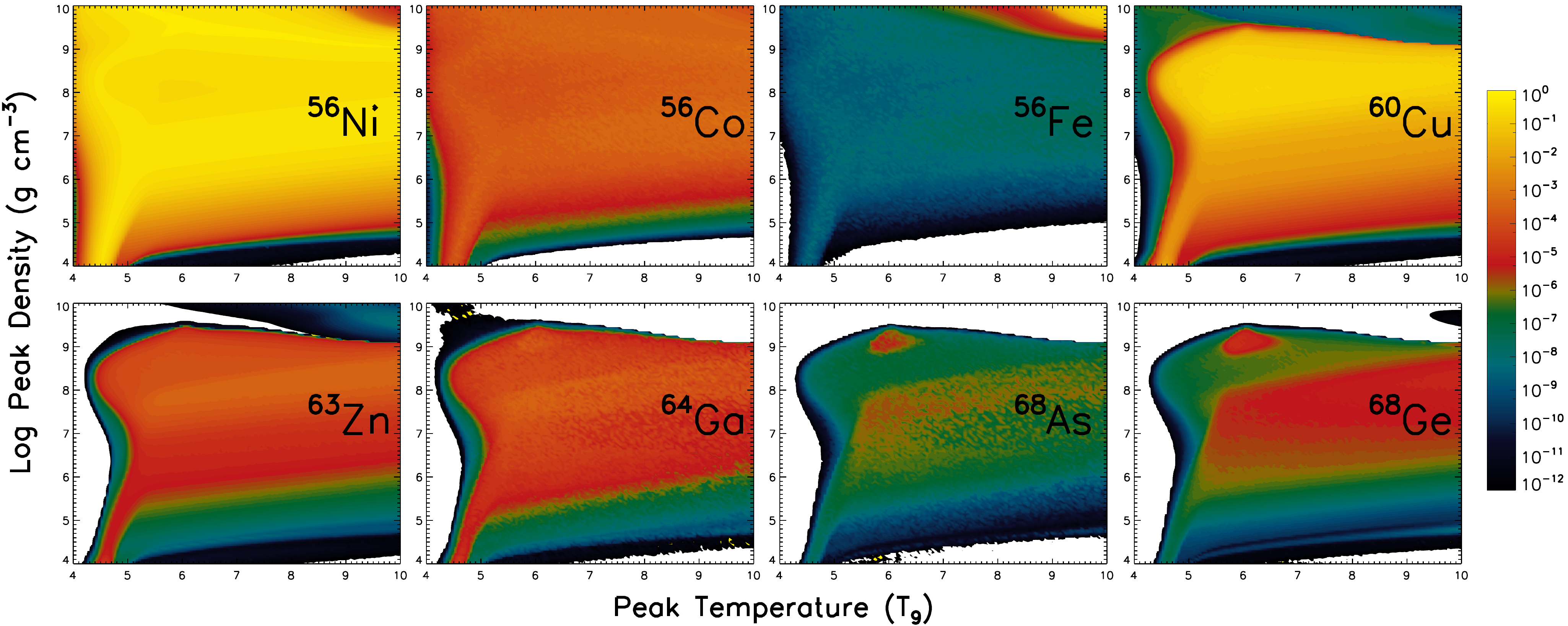}
\caption{Final mass fractions of isotopes with protons and neutrons
near the magic number 28 (second family) in the peak
temperature--density plane for the power-law profile at $Y_{e} =
0.52$. The white colored space corresponds to values below the color
scale shown. From left to right, the first row corresponds to
\ux{56}{Ni}, \ux{56}{Co}, \ux{56}{Fe} and \ux{60}{Cu}, and the
second row corresponds to \ux{63}{Zn}, \ux{64}{Ga}, \ux{68}{As} and
\ux{68}{Ge}.} \label{fig:contour_PL2_ye0520_ni56-like}
\end{figure}

The types of freeze-out discussed so far (normal, $\alpha$-rich,
$\alpha$$p$-rich, and the chasm) tend to favor the production of
nuclei with proton and neutron numbers in the locality of the magic
number 28. Figure \ref{fig:contour_PL2_ye0520_ni56-like} shows a
sample of such nuclei, and Table \ref{tab:ni56-like_yields} provides
the complete list. These isotopes tend to dominate the final
composition for most initial electron fraction values. The final
mass fractions in Figure \ref{fig:contour_PL2_ye0520_ni56-like}
demonstrate homogeneous structures within the temperature--density
plane, implying that these isotopes do not sustain any EST during
the evolution. The restriction of the remnant QSE cluster and the
accumulation of nuclear flow among these isotopes are responsible
for the absence of ESTs. The accumulation of flow stems from the
fact that nuclei with proton or neutron numbers near the magic
number 28 tend to maximize their binding energy per nucleon. As a
result, such nuclei are relatively more bound compared to nuclei
with nucleon numbers far from the magic number values, and their
production within a network of reactions is favored. We classify
isotopes which do not sustain any EST during freeze-out expansions
and tend to dominate the final composition into a ``second family''
of isotopes. We have demonstrated that nuclei whose neutron or
proton number is near the magic number 28 belong to the second
family. Below, we show that nuclei with neutron numbers near the
magic numbers 50 and 82 also belong to the second family.

\begin{figure}[htp]
\includegraphics[width=0.95\textwidth]{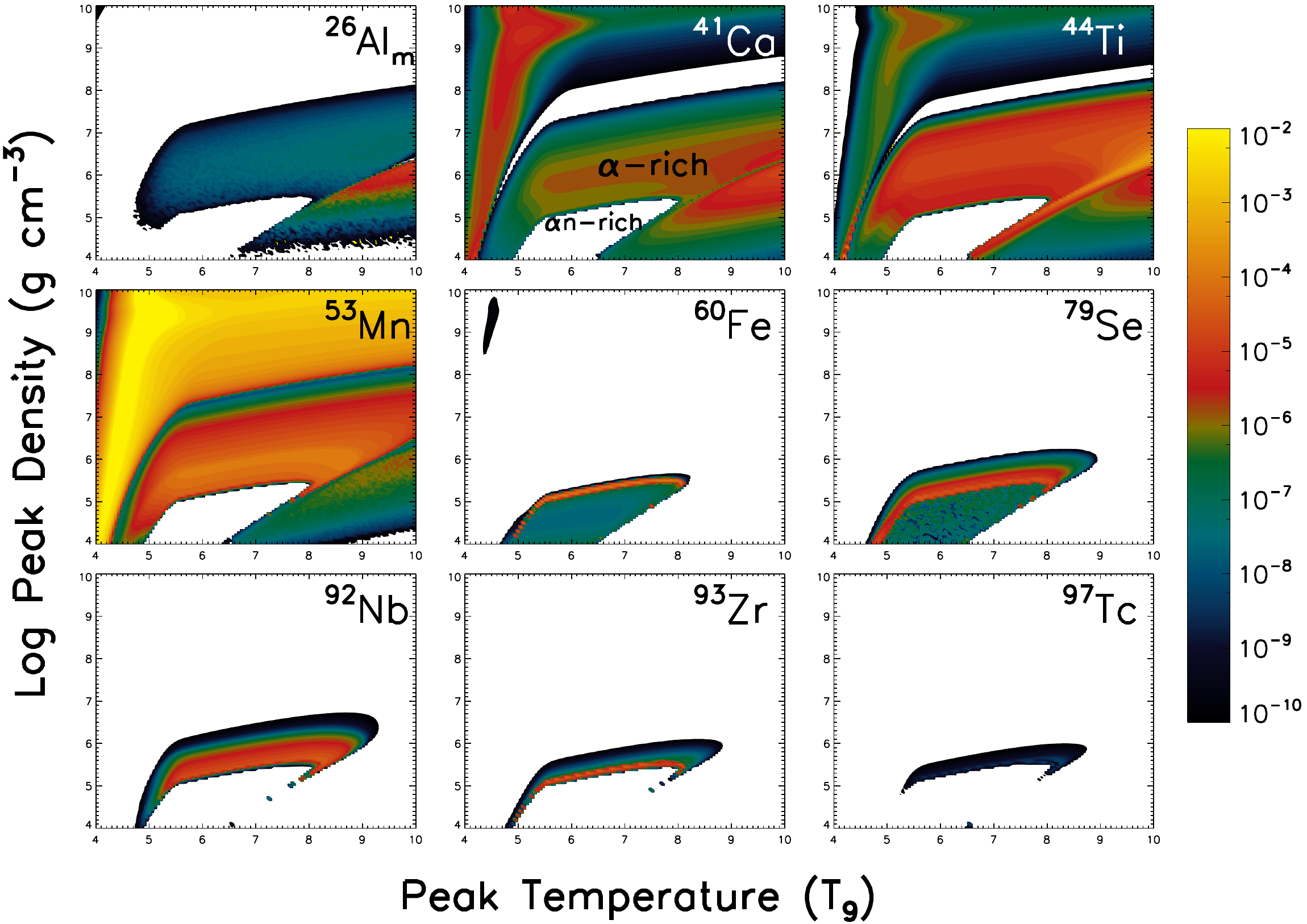}
\caption{Final mass fractions of select radioactivities in the peak
temperature–-density plane for the exponential profile at $Y_{e} =
0.48$. The white colored space corresponds to values below the color
scale shown. From left to right, the first row corresponds to
\ux{26}{Al}\mr{_m}, \ux{41}{Ca}, and \ux{44}{Ti}, the second row
corresponds to \ux{53}{Mn}, \ux{60}{Fe}, and \ux{79}{Se}, and the
third row corresponds to \ux{92}{Nb}, \ux{93}{Zr}, and \ux{97}{Tc}.
The $\alpha$-rich and $\alpha$$n$-rich freeze-out regions are
labeled on the temperature--density plane for \ux{41}{Ca}.}
\label{fig:contour_AD1_ye0480_traces}
\end{figure}

\begin{figure}[htp]
\includegraphics[width=0.95\textwidth]{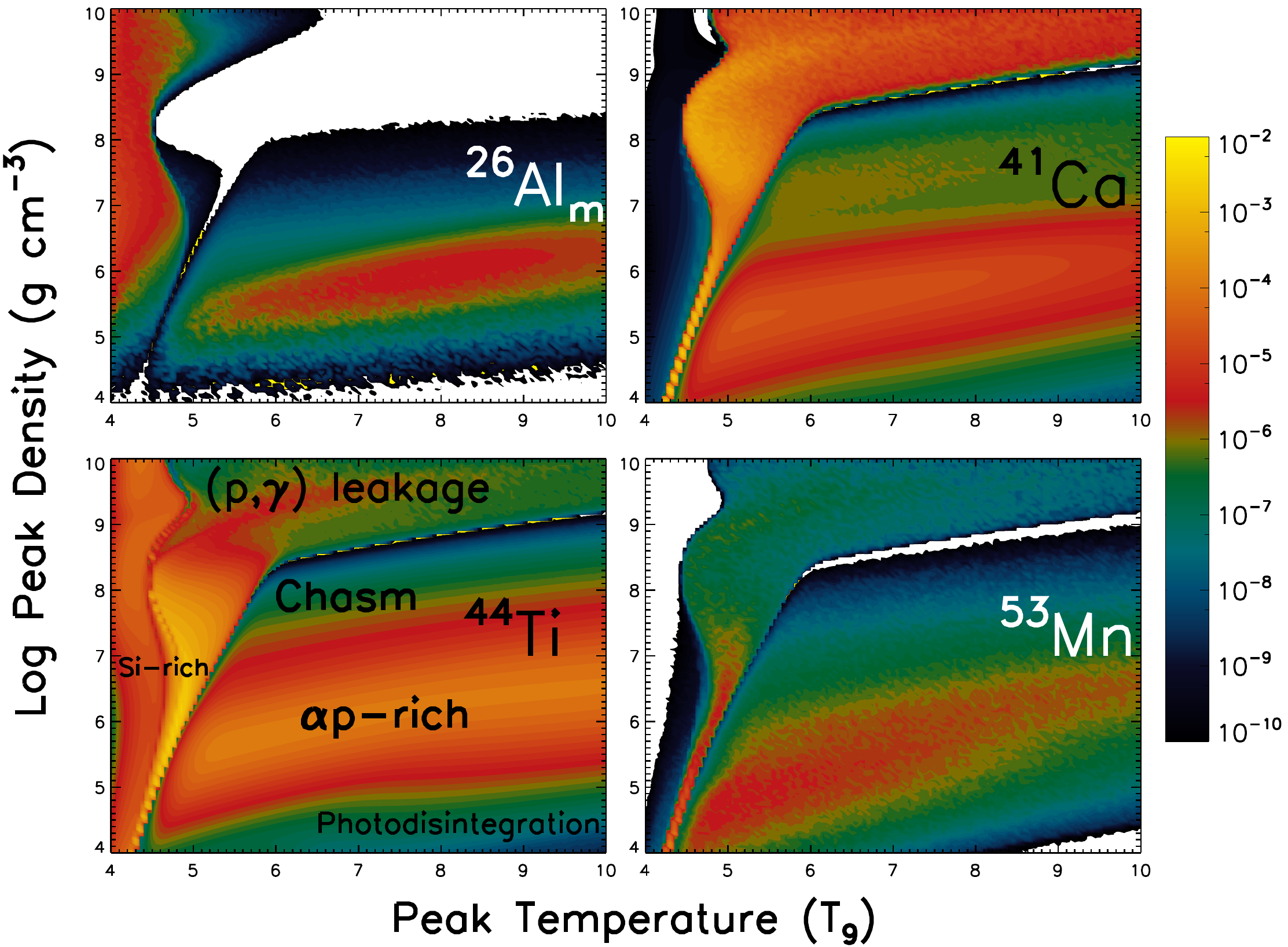}
\caption{Final mass fractions of select radioactivities in the peak
temperature–-density plane for the exponential profile at $Y_{e} =
0.52$. The white colored space corresponds to values below the color
scale shown. From left to right, the first row corresponds to
\ux{26}{Al}\mr{_m}, and \ux{41}{Ca}, and the second row corresponds
to \ux{44}{Ti} and \ux{53}{Mn}.}
\label{fig:contour_AD1_ye0520_traces}
\end{figure}

Figures \ref{fig:contour_AD1_ye0480_traces} and
\ref{fig:contour_AD1_ye0520_traces} show the temperature--density
planes of select radioactivities up to mass $A=97$ which have
non-negligible yields for the corresponding initial $Y_e$ values.
The regions of the $\alpha$$n$-rich freeze-out and
\mr{(p,\gamma)}-leakage regime are labeled. These two types of
freeze-out expansions are not manifested for initially symmetric
matter. The temperature--density planes depict a region-divided
structure, indicating that these radioactivities belong to the first
family of isotopes. The decay timescale of \ux{26}{Al}\mr{_m} is
approximately 2 s. Consequently, yield values are seen only for the
exponential profile where the expansion timescale is less than a
second, while for the power-law it decays prior to complete
freeze-out. On the contrary, \ux{26}{Al}\mr{_g} is mostly produced
during the power-law expansion for $Y_{e}\geqslant0.5$. \ux{41}{Ca}
is produced during the $\alpha$-rich and $\alpha$$p$-rich
freeze-outs for the full range of our initial electron fraction
values, and also during the \mr{(p,\gamma)}-leakage regime (Figure
\ref{fig:contour_AD1_ye0520_traces}) for the exponential profile and
$Y_{e}>0.5$. The \mr{(\alpha,\gamma)} and \mr{(\alpha,p)} channels
control its production for $Y_{e}=0.48$, while the \mr{(p,\gamma)}
and the weak reactions impact its synthesis for $Y_{e}\geqslant0.5$.
In addition, the \mr{(\alpha,p)} channels shape the locus of the
borderline between the \mr{(p,\gamma)}-leakage and Si-rich regimes
in the contour plot for $Y_{e}>0.5$. \ux{53}{Mn} is produced mostly
during the normal freeze-out for $Y_{e}\leqslant0.5$, although it
has significant yields from the $\alpha$-rich and $\alpha$$p$-rich
freeze-outs for the power-law expansion. It is relatively
insensitive to reaction rates for $Y_{e}<0.5$, while the
\mr{(p,\gamma)} and weak reactions have significant impact for
$Y_{e}\geqslant0.5$. The \mr{(\alpha,p)} and \mr{(p,\gamma)}
channels affect the borderline between the \mr{(p,\gamma)}-leakage
and Si-rich regimes.

Radioactivities heavier than mass $A=60$ are produced primarily in
neutron-rich environments for the exponential and power-law
profiles. Specifically, they may either be produced during an
$\alpha$$n$-rich freeze-out \citep{woosley_1992_aa} or by a process
that combines features between the $\alpha$-rich and
$\alpha$$n$-rich freeze-outs (Figure
\ref{fig:contour_AD1_ye0480_traces}). The $\alpha$$n$-rich
freeze-out occurs for relatively low initial electron fraction
values and its locus is constrained to regions of low peak densities
in the contour plots. The combination of high peak temperatures and
low peak densities allows the establishment of a photodisintegration
regime early in the evolution. The balance between the \pen\ and
\nep\ reactions maintains the electron fraction values below 0.5,
which favor an overproduction of neutrons against protons. Such
values for $Y_e$ allow the major nuclear flows to bypass the doubly
magic nucleus \ux{56}{Ni} and heavier elements are produced
\citep{hartmann_1985_aa,woosley_1992_aa,magkotsios_2010_aa}. The QSE
cluster shifts upward in mass, but it is not localized solely around
the magic number 28. Instead, neutron capture reactions shift the
cluster to heavier masses and pile up nuclear flow in the locality
of nuclei with neutron magic numbers 50 and 82. The concentration of
nuclear flow around these nuclei maintains the equilibrium structure
in their locality and prevents them from sustaining ESTs. Once
again, the flow concentration near nuclei with magic numbers 50 and
82 is a nuclear structure effect. These nuclei maximize locally the
nuclear binding energy per nucleon and are relatively more bound
compared to other nuclei with nucleon numbers away from the magic
number series. Consequently, isotopes such as \ux{86}{Kr},
\ux{87}{Rb}, \ux{88}{Sr}, and \ux{122}{Zr} belong to the second
family of isotopes and dominate the final composition along with
free $\alpha$-particles and neutrons. During the evolution, the
excess of free neutrons guarantees a large-scale equilibrium
structure maintained primarily by chains of \mr{(n,\gamma)} and
\mr{(p,n)} reactions in equilibrium. Isotopes of the first family
with mass $A\gtrsim60$ including the radioactivities \ux{60}{Fe} and
\ux{79}{Se} sustain an EST when \mr{(p,n)} reactions in their
locality break equilibrium.

\begin{figure}[htp]
\includegraphics[width=0.95\textwidth]{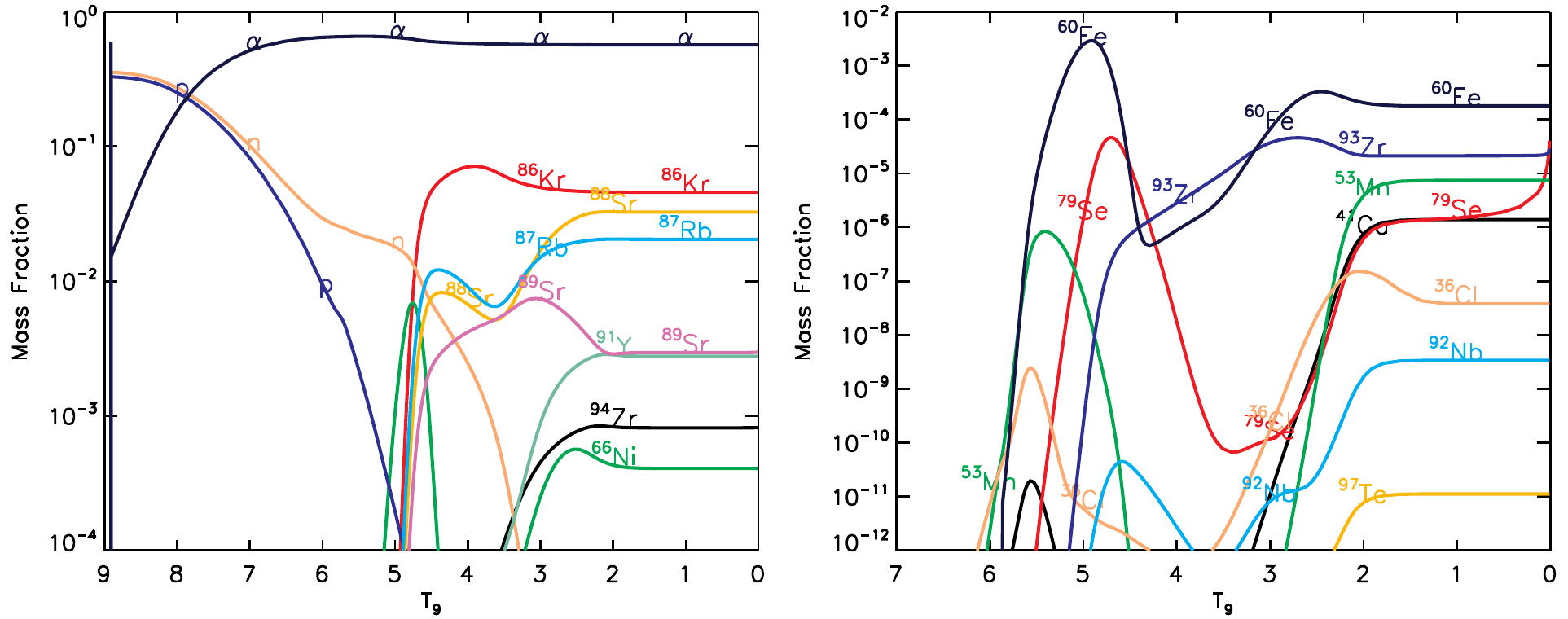}
\caption{Mass fraction evolution of dominant yields (left panel) and
radioactivities (right panel) for the transition region between the
$\alpha$-rich and $\alpha$$n$-rich freeze-outs. Initial conditions
are $T_{9}=9$, $\rho = 5\times 10^7$ g cm$^{-3}$, and $Y_{e} = 0.48$
for the power-law profile (3304 isotope network).}
\label{fig:mass_PL2_ye0480_dom_traces}
\end{figure}

\begin{figure}[htp]
\includegraphics[height=0.8\textheight]{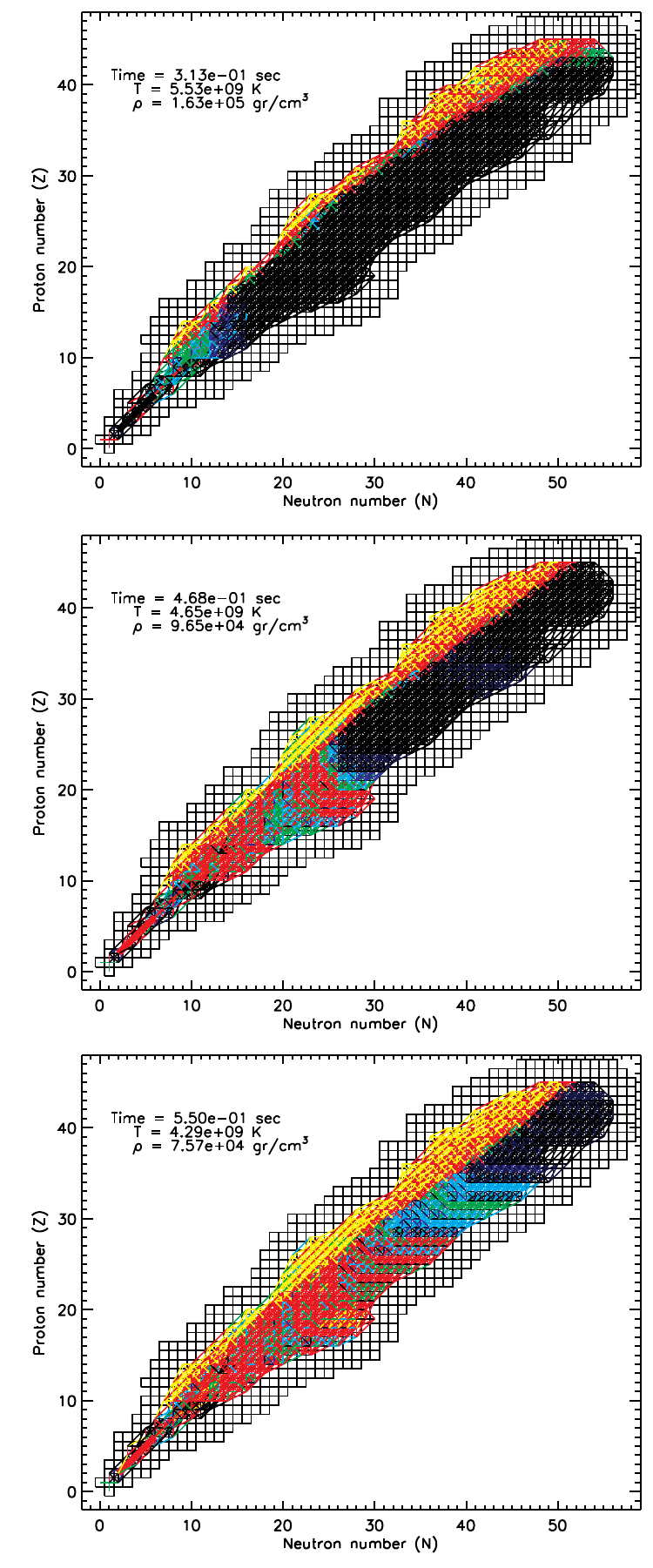}
\caption{QSE cluster motion in the chart of nuclides for the
transition region between $\alpha$-rich and $\alpha$$n$-rich
freeze-outs. The QSE cluster remnant bypasses the isotopes of the
second family near the magic number 28, and condenses around the
corresponding isotopes near the magic number 50. Each colored line
corresponds to a nuclear reaction and indicates the level of nuclear
flow transferred between the isotopes connected. Normalized flows
$\phi$ are colored black for $0 \le \phi < 0.01$, navy for $0.01 \le
\phi < 0.05$, blue for $0.05 \le \phi < 0.1$, cyan for $0.1 \le \phi
< 0.4$, green for $0.4 \le \phi < 0.8$, red for $0.8 \le \phi <
1.0$, and yellow for $\phi$=1.0. Small $\phi$ values indicate
reactions in equilibrium, while $\phi$=1.0 implies pure one-way
nuclear flow transfer. } \label{fig:charts_an-rich}
\end{figure}

The production of radioactivities such as \ux{92}{Nb}, \ux{93}{Zr},
and \ux{97}{Tc} by the exponential and power-law expansions is
favored only within the narrow transition region between the
$\alpha$-rich and $\alpha$$n$-rich freeze-outs. Figure
\ref{fig:mass_PL2_ye0480_dom_traces} shows the mass fraction
evolution of dominant elements and radioactivities within this
region. The free neutrons are depleted below the NSE threshold and
do not allow the neutron capture reactions to shift the QSE cluster
until the magic number 82. The flows are blocked around the magic
number 50. This effect is illustrated in Figure
\ref{fig:charts_an-rich}. The yields of \ux{86}{Kr}, \ux{87}{Rb},
and \ux{88}{Sr} still dominate the final composition, but are
slightly enhanced compared to the region of the $\alpha$$n$-rich
freeze-out. The constraint of nuclear flow near \ux{86}{Kr}
maximizes the yields of \ux{60}{Fe} and \ux{79}{Se}, and allows the
production of \ux{92}{Nb}, \ux{93}{Zr}, and \ux{97}{Tc} (Figure
\ref{fig:contour_AD1_ye0480_traces}).

\section{REACTION RATE SENSITIVITY STUDY FOR SELECT RADIOACTIVITIES}
\label{sec:reactions}

We perform a sensitivity study on reaction rates related to the
synthesis of radioactivities for the exponential and power-law
expansions, to identify the reactions which are primarily
responsible for their production in core-collapse supernovae. These
critical reactions determine whether an EST takes place or not.
Reaction channels and individual rates are either multiplied by a
factor or removed from the network. Strong reaction rates are
multiplied by factors of 100 or 0.01, while the corresponding
factors for the weak reactions are 1000 or 0.001. These factors are
adequate to facilitate the identification of trends in the yields,
although they exceed experimental uncertainties in most cases.
Sensitivity studies where reactions were varied within experimental
uncertainty ranges have been performed by \citet{hoffman_2010_aa}
and \citet{tur_2010_aa}. Our calculations include rates for weak
interactions
\citep{fuller_1980_aa,fuller_1982_aa,fuller_1982_ab,oda_1994_aa,
langanke_2001_aa}, the theoretical rates of \citet{rath_2000_aa},
and select experimental rates for capture and photodisintegration
reactions.

\begin{figure}[htp]
\centering\includegraphics[width=0.9\textwidth]{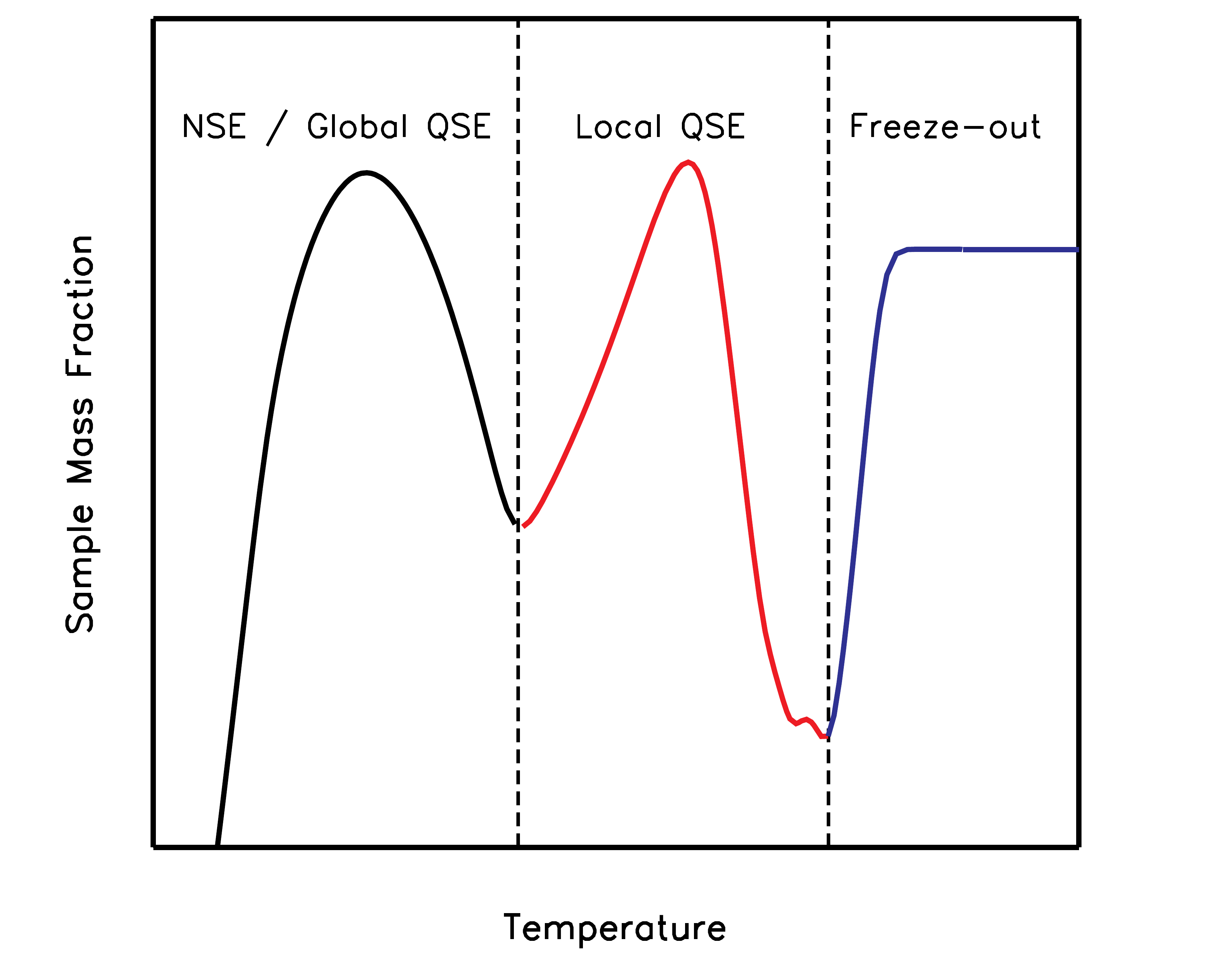}
\caption{Mass fraction schematic for decreasing temperature, where
the arc structure is illustrated. Such mass fraction profiles are
representative of the chasm, $\alpha$-rich, $\alpha$$p$-rich, and
$\alpha$$n$-rich freeze-outs. The first arc (black) describes the
mass fraction trend during a large-scale QSE state. The second arc
(red) describes the mass fraction trend once the isotope is outside
the large equilibrium cluster, and its trends may be explained by
local equilibrium states (local QSE). The blue ascending track is
denoted as third arc, and is related to a mixture of local QSE and
non-equilibrium nucleosynthesis. Mass fractions during an
$\alpha$-rich freeze-out have only an ascending track past the first
arc, while for the chasm region the mass fractions have only one
arc.} \label{fig:arc_cartoon}
\end{figure}

The structure of the temperature--density plane for isotopes of the
first family is affected by certain key reactions per isotope, in
combination with the expansion timescale. Specific reactions such as
the 3$\alpha$, $\alpha\alpha n$, \pen, \nep, and combinations of a
large number of weak reactions impact all mass fractions
simultaneously either by transferring nuclear flow to the QSE
cluster, or by contributing to electron fraction variations
\citep{fuller_1995_aa,mclaughlin_1995_aa,mclaughlin_1996_aa,
aprahamian_2005_aa,surman_2005_aa,liebendorfer_2008_aa}. However,
the local equilibrium patterns are controlled by reactions in the
locality of each isotope. Table \ref{tab:nuclear_reactions} lists
the reactions that impact the synthesis of radioactivities in the
mass range $12\leqslant A\leqslant122$ produced during freeze-out
expansions. In Table \ref{tab:nuclear_reactions} the contribution of
a reaction is focused on specific parts of the isotope's mass
fraction curve. Below we use the term ``arc'' frequently, so it is
convenient to provide a visualization of this structure. Figure
\ref{fig:arc_cartoon} shows a sample mass fraction evolution of an
isotope for decreasing temperature. Two local minimum points of the
mass fraction curve are identified. These points separate the curve
in three parts (arcs). The first part (black arc) is related to the
mass fraction evolution when the isotope participates in a
large-scale QSE cluster (global QSE). The second and third parts
(red and blue arcs respectively) are related to mass fraction trends
when the isotope either participates in small-scale equilibrium
clusters (local QSE) or does not belong to any cluster at all
(non-equilibrium nucleosynthesis). Since the third arc is not a full
arc, it is also mentioned as ``ascending track''. Note that the mass
fraction curve may be limited to two arcs, i.e. the first arc and
the ascending track until freeze-out (for instance, see Figure
\ref{fig:mass_PL2_ye0480_dom_traces}).

The \ux{26}{Al} yield has an average value $X(\ux{26}{Al})\sim
10^{-7}$ within the $\alpha$-rich and $\alpha$$p$-rich freeze-out
regimes for the exponential profile only, where the metastable state
\ux{26}{Al}\mr{_m} has significant yield (Figures
\ref{fig:contour_AD1_ye0480_traces} and
\ref{fig:contour_AD1_ye0520_traces}).
\mr{\ux{26}{Al}_g(p,\gamma)\ux{27}{Si}} and
\mr{\ux{26}{Al}_g(\alpha,p)\ux{29}{Si}} configure the chasm features
and the characteristic arcs in the mass fraction profiles during the
$\alpha$-rich and $\alpha$$p$-rich freeze-outs. They also control
the flow between the $N=13$ and $N=15$ isotones. The equilibrium
break of \mr{\ux{26}{Al}_g(\alpha,p)\ux{29}{Si}} marks the
appearance and controls the depth of the chasm in the contour plot
and the first dip in the \ux{26}{Al} mass fraction profile. The
equilibrium break of \mr{\ux{26}{Al}_g(p,\gamma)\ux{27}{Si}} during
the $\alpha$$p$-rich freeze-out transfers flow from \ux{26}{Al} to
heavier isotopes along the $N=13$ isotone and configures the shape
and dip of the second arc in the \ux{26}{Al} mass fraction.
Additional reactions that contribute similarly are
\mr{\ux{25}{Mg}(p,\gamma)\ux{26}{Al}_g},
\mr{\ux{25}{Al}(p,\gamma)\ux{26}{Si}},
\mr{\ux{26}{Si}(p,\gamma)\ux{27}{P}}, and
\mr{\ux{27}{P}(p,\gamma)\ux{28}{S}}. The
\mr{\ux{24}{Mg}(\alpha,\gamma)\ux{28}{Si}} reaction distributes flow
within the silicon group and shapes the ascending track to the
\ux{26}{Al} mass fraction past the arcs during the $\alpha$$p$-rich
freeze-out. The yield of \ux{26}{Al} is largely dependent on the
collective flow transfer by weak reactions toward symmetric nuclei.
Specific weak reactions with the largest contribution for
\ux{26}{Al} are listed in Table \ref{tab:nuclear_reactions}.

\ux{41}{Ca} is an example of composite contribution from
\mr{(p,\gamma)} reactions along its own and neighboring isotone
lines which are connected by \mr{(\alpha,\gamma)} reactions. Table
\ref{tab:nuclear_reactions} lists the related reactions for
$Y_{e}\geqslant0.5$, and reactions that transfer flow for
neutron-rich compositions. For proton-rich environments \ux{41}{Ca}
is also dependent on the collective flow transfer by weak reactions.

Radioactivities with mass $A\gtrsim60$ depend mostly on neutron
captures and weak reactions. The yield of \ux{53}{Mn} depends only
on the collective flow transfer by weak reactions. The weak
reactions with the largest contribution are listed in Table
\ref{tab:nuclear_reactions}. \mr{\ux{86}{Kr}(\alpha,n)\ux{89}{Sr}}
is the main flow distributor within the small cluster in the
locality of the neutron magic number $N=50$ and it affects the mass
fractions of \ux{60}{Fe}, \ux{79}{Se}, \ux{92}{Nb}, \ux{93}{Zr}, and
\ux{97}{Tc}. \mr{\ux{60}{Fe}(p,n)\ux{60}{Co}} shapes the second arc
of the \ux{60}{Fe} mass fraction, and its strength determines the
degree of the yield's depletion.
\mr{\ux{60}{Fe}(\alpha,n)\ux{63}{Ni}} contributes to the formation
of the arc, while \mr{\ux{59}{Fe}(n,\gamma)\ux{60}{Fe}} and
\mr{\ux{58}{Fe}(\alpha,n)\ux{61}{Ni}} regulate the arc's amplitude
and slope respectively. Most notably,
\mr{\ux{60}{Fe}(n,\gamma)\ux{61}{Fe}} does not impact the
\ux{60}{Fe} mass fraction and yield for the freeze-out expansions.
The mass fraction profile of \ux{79}{Se} is marked by a sharp
ascending track at complete freeze-out which increases the yield by
an order of magnitude. The abrupt flow transfer stems mostly from
the collective contribution of the weak reactions, with major
contribution from \mr{\ux{79}{As}(,e^{-}\;\nu_{e})\ux{79}{Se}}.
Additional reactions that contribute to this ascending track prior
to freeze-out are \mr{\ux{77}{Se}(n,\gamma)\ux{78}{Se}} and
\mr{\ux{78}{Se}(n,\gamma)\ux{79}{Se}}. Its mass fraction arc is
formed by \mr{\ux{79}{As}(p,n)\ux{79}{Se}}. Similarly, the mass
fraction arc for \ux{92}{Nb} is formed by
\mr{\ux{92}{Zr}(p,n)\ux{92}{Nb}}, and the arc for \ux{97}{Tc} by
\mr{\ux{97}{Mo}(p,n)\ux{97}{Tc}},
\mr{\ux{97}{Tc}(n,\gamma)\ux{98}{Tc}}, and
\mr{\ux{98}{Mo}(p,n)\ux{98}{Tc}}. \ux{93}{Zr} is in the locality of
the neutron magic number $N=50$, and its mass fraction profile
monotonically increases up to $X(\ux{93}{Zr})\sim 10^{-4}$ in the
transition region between the $\alpha$-rich and $\alpha$$n$-rich
freeze-outs. The rate strength of
\mr{\ux{92}{Zr}(n,\gamma)\ux{93}{Zr}} shapes the yield value by
transferring flow to \ux{93}{Zr} from the isotopes with neutron
number $N=50$.

\section{YIELDS FROM NON-MONOTONIC PROFILES}
\label{sec:non-monotonic}

The exponential and power-law profiles discussed so far involve the
initial formation of a large-scale QSE cluster and subsequent states
of small-scale clusters. Non-equilibrium nucleosynthesis appears
only during the freeze-out stage at the end of the evolution. The
non-monotonic expansion profiles (Section \ref{sec:profiles}) may
have non-equilibrium intervals followed by the formation of a
large-scale QSE cluster, resulting in final compositions which
cannot be achieved by monotonic profiles. Table
\ref{tab:profile_comparison} lists the differences between the
monotonic and non-monotonic profiles. The keywords related to the
evolution of the QSE cluster are ``hierarchical'' and ``periodic''
for the monotonic and non-monotonic profiles respectively. The term
hierarchical denotes the gradual dissolution of the single
large-scale QSE cluster to multiple small-scale clusters, while the
term periodic describes the sequence of transitions from NSE and
global QSE, to local QSE and non-equilibrium phase, to global QSE
again and then local QSE and non-equilibrium phase until freeze-out.

The existence of ESTs during non-monotonic profiles depends on the
combination of the extremum point values for the temperature and
density. \citet{arcones_2007_aa} report ranges of $0.4\leqslant
T_9\leqslant2$ and $10^2\leqslant\rho\leqslant10^4$ g cm$^{-3}$ for
the temperature and density behind the reverse shock, while
\citet{wanajo_2011_aa} report a temperature range of $1.5\leqslant
T_9\leqslant3$. The flow transfer by reactions out of equilibrium
among scattered small-scale equilibrium clusters impacts the mass
fraction evolution dramatically. The flow patterns are very
sensitive to variations in the values of the temperature and density
extremum points. This sensitivity diversifies the production
mechanisms significantly. Table \ref{tab:non-monotonic_yields} lists
the combinations of temperature and density extremum points within
our data set which tend to maximize the yields of radioactivities.
The peak temperature for the non-monotonic expansions is chosen to
be $T_9=9$, and the peak density and initial electron fraction
values range between $10^6\leqslant\rho\leqslant10^8$ g cm$^{-3}$
and $0.48\leqslant Y_e \leqslant 0.52$ respectively. These peak
temperature and density values are large enough to establish NSE
early in the evolution, so that the initial composition dependence
is removed. Below we present the details of certain profiles which
produce simultaneously most of the radioactivities in the mass range
$12\leqslant A\leqslant122$. These radioactive isotopes could be
transferred to upper (cooler) mass layers by mixing processes during
the explosion, where the effective lifetime of the radioactivities
is longer \citep{tur_2010_aa}, and possibly released to the
interstellar medium.

\begin{figure}[htp]
\includegraphics[width=0.95\textwidth]{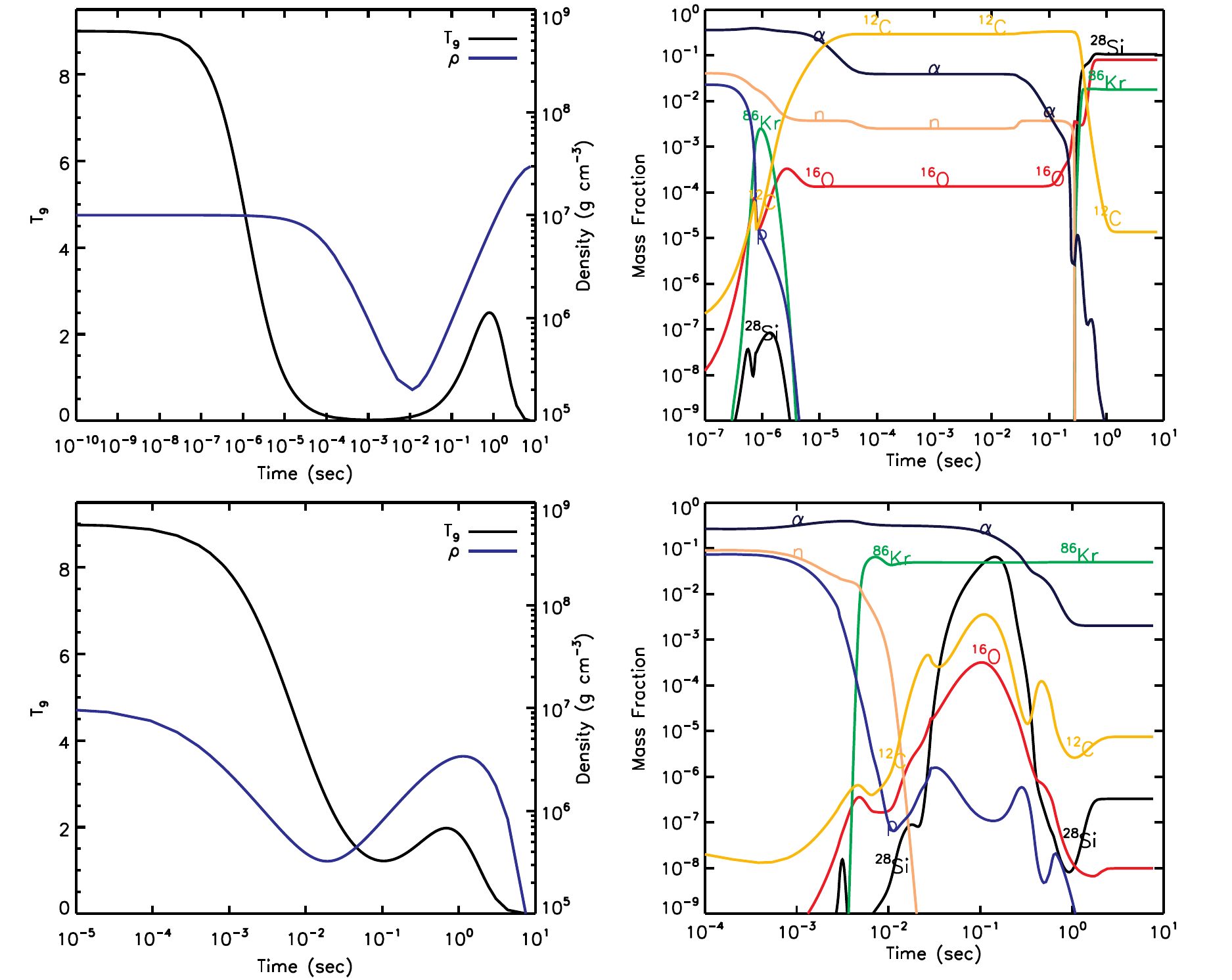}
\caption{Non-monotonic temperature and density trajectories (left
column) and key mass fraction profiles (right column) for initial
conditions $T_{9}=9$, $\rho = 10^7$ g cm$^{-3}$, and $Y_{e} = 0.48$.
The top row is a profile where the \ux{60}{Fe} yield is
approximately maximized, while the profile at the bottom row tends
to maximize the \ux{44}{Ti} yield within our data set.}
\label{fig:MC_ye0480_fe60_ti44}
\end{figure}

Figure \ref{fig:MC_ye0480_fe60_ti44} shows the thermodynamic and
mass fraction evolutions of two non-monotonic profiles for initially
neutron-rich composition ($Y_e=0.48$). The extremum values for the
profile of the top row are $T_9^{min}=0.02$, $T_9^{max}=2.5$,
$\rho_{min}=2\times10^5$ g cm$^{-3}$, and $\rho_{max}=3\times10^7$ g
cm$^{-3}$. This profile approximately maximizes the \ux{60}{Fe}
yield within our data set. During the initial temperature decrement
(until $t=10^{-4}$ s) the density is relatively fixed to its peak
value. The conditions at the $T_9=5$ NSE threshold are similar to an
$\alpha$-rich freeze-out in neutron-rich matter, where the 3$\alpha$
forward rate dominates its inverse photodisintegration and the
protons are rapidly consumed. However, the temperature evolution is
fast enough that it prevents significant flow transfer beyond
\ux{16}{O}, and the QSE cluster dissolves without shifting upward in
mass. The temperature nearly reaches freeze-out levels after
$t=10^{-4}$ s, and \ux{12}{C} dominates the composition, with
significant mass fraction values for free neutrons,
$\alpha$-particles, and \ux{16}{O}. The density decrease from
$t=10^{-4}$ s until $t=10^{-2}$ s does not impact the mass fractions
due to low temperature values. The subsequent temperature and
density increase result in carbon burning primarily by the
\mr{\ux{12}{C}(\ux{12}{C},p)\ux{23}{Na}} and
\mr{\ux{12}{C}(\ux{12}{C},\alpha)\ux{20}{Ne}} reactions. A second
large-scale QSE cluster is formed and the low electron fraction
values in combination with the free neutron abundance guarantee flow
transfer near \ux{86}{Kr} through \mr{(n,\gamma)} and \mr{(p,n)}
reactions. The temperature maximum is slightly lower than the
typical silicon burning threshold $T_9\sim3$ \citep{iliadis_2007_aa}
at $t=1$ s, and the final composition is dominated mostly by carbon
burning products and elements in the locality of the $N=50$ neutron
magic number, while elements near the $N=28$ magic number are
severely underproduced. The final composition includes significant
yields for the radioactivities listed in Table
\ref{tab:non-monotonic_yields}, excluding \ux{44}{Ti} and
\ux{97}{Tc}.

The second row of Figure \ref{fig:MC_ye0480_fe60_ti44} corresponds
to a profile with extremum values $T_9^{min}=1.4$, $T_9^{max}=1.94$,
$\rho_{min}=3.3\times10^5$ g cm$^{-3}$, and
$\rho_{max}=3.4\times10^6$ g cm$^{-3}$. This profile approximately
maximizes the \ux{44}{Ti} yield within our data set. The flow
transfer by the 3$\alpha$ rate to the QSE cluster occurs over a
longer interval compared to the profile of the top row in Figure
\ref{fig:MC_ye0480_fe60_ti44}. The QSE cluster shifts upward in mass
until the $N=50$ neutron magic number, producing \ux{86}{Kr} and the
radioactivities \ux{92}{Nb}, \ux{93}{Zr} and \ux{97}{Tc}. Up to the
point where the \ux{86}{Kr} abundance is maximized (near $t=10^{-2}$
s), the process resembles the evolution of exponential and power-law
profiles in the transition regime between the $\alpha$-rich and
$\alpha$$n$-rich freeze-outs (see Section \ref{sec:monotonic} and
Figure \ref{fig:mass_PL2_ye0480_dom_traces}). The first
$\alpha$-capture reactions to break equilibrium appear in the mass
region of \ux{12}{C} and \ux{16}{O}. The increasing number of
$\alpha$-capture reactions which break equilibrium within this mass
region and the silicon group shift the low mass border of the QSE
cluster toward heavier nuclei. The $\alpha$-capture reactions with
the largest net flows are blocked within the silicon group, because
the Coulomb repulsion for the specific thermodynamic conditions
prevents the $\alpha$-capture reactions involving heavier nuclei
from acquiring large flow values. However, the QSE cluster continues
to shift upward in mass. The blockage of the largest flows for the
$\alpha$-captures is a non-equilibrium effect, and the precise
nuclear mass range where these large flows are localized in depends
on the details of the specific expansion profile. For the profile of
the bottom row in Figure \ref{fig:MC_ye0480_fe60_ti44} the largest
flows shift in mass until the \ux{40}{Ca} -- \ux{44}{Ti} --
\ux{48}{Cr} region and produce these isotopes with a mass fraction
$X\sim10^{-3}$. The largest flows shift downward in mass next, while
the temperature approaches its minimum value at time $t=10^{-1}$ s.
At this time, the \ux{28}{Si} mass fraction reaches its maximum
value (bottom right panel in Figure \ref{fig:MC_ye0480_fe60_ti44}).
The subsequent temperature and density increase until time $t=1$ s
relocates the largest flows for the $\alpha$-captures back to the
\ux{40}{Ca} -- \ux{44}{Ti} -- \ux{48}{Cr} region at the cost of the
\ux{28}{Si} abundance. \ux{44}{Ti} dominates the final composition
with a yield $X(\ux{44}{Ti})=0.286$. Other isotopes to be produced
in significant amounts include \ux{41}{Ca} and \ux{53}{Mn}. Isotopes
near the mass $A=56$ are underproduced, since their equilibrium
abundances are not favored by the QSE formation, and the largest
flows of $\alpha$-capture reactions never reach this mass region.

The non-monotonic profiles of Figure \ref{fig:MC_ye0480_fe60_ti44}
demonstrate cases of final compositions which cannot be achieved by
monotonic profiles, such as the exponential and power law profiles.
Both non-monotonic expansions include significant intervals of
non-equilibrium nucleosynthesis followed by the reformation of
equilibrium clusters. This feature results in a final composition
dominated by silicon group elements which belong to the first family
of isotopes, and neutron-rich isotopes of the second family near the
mass $A=86$. Isotopes of the second family near the mass $A=56$ are
underproduced. During monotonic expansions, the QSE cluster size
decreases gradually during the evolution and the non-equilibrium
part of nucleosynthesis is always constrained near freeze-out. This
feature results in the dominance of isotopes in the second family
only, and does not allow patterns such as those of Figure
\ref{fig:MC_ye0480_fe60_ti44}.

\begin{figure}[htp]
\includegraphics[width=0.95\textwidth]{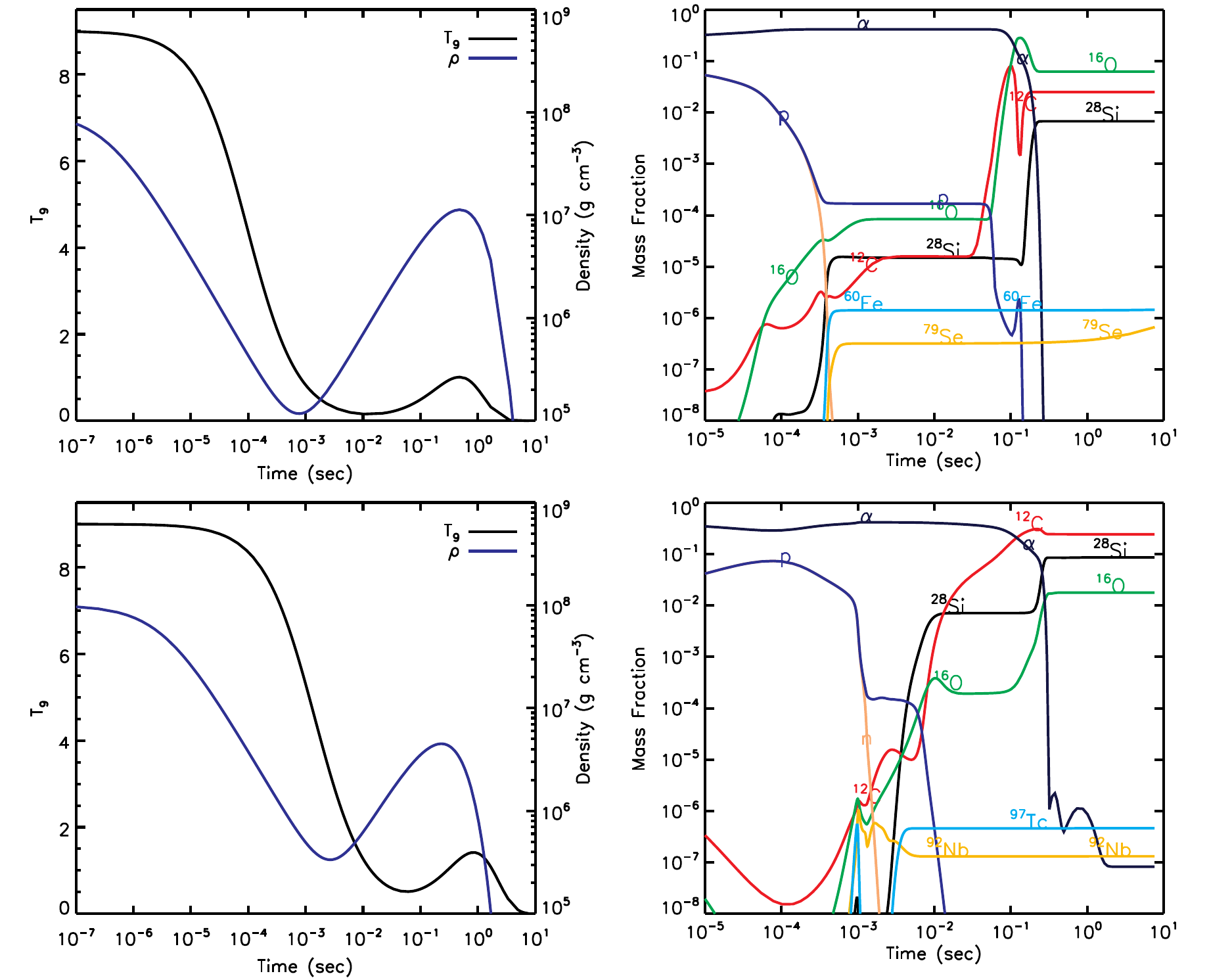}
\caption{Non-monotonic temperature and density trajectories (left
column) and key mass fraction profiles (right column) for initial
conditions $T_{9}=9$, $\rho = 10^8$ g cm$^{-3}$, and $Y_{e} = 0.50$.
The top row is a profile where the \ux{60}{Fe} yield is
approximately maximized, while the profile at the bottom row tends
to maximize the \ux{97}{Tc} yield within our data set.}
\label{fig:MC_ye0500_fe60_tc97}
\end{figure}

The mixing processes during the supernova explosion impact
significantly the nucleosynthesis mechanisms. The majority of
one-dimensional supernova models tend to position the mass-cut near
the region where the electron fraction begins to decrease below
$Y_e=0.5$ \citep{woosley_1973_aa,weaver_1993_aa,woosley_1995_aa,
thielemann_1996_aa,rauscher_2002_aa,woosley_2002_aa,limongi_2003_aa,chieffi_2004_aa}.
As a result, the subsequent supernova shock wave in these models
traverses material which has initially symmetric or nearly symmetric
composition. Our exponential and power-law analysis has demonstrated
that the radioactivities for $A\geqslant60$ are not produced in
significant amounts during explosions within symmetric matter which
lack mixing processes. However, there exist types of non-monotonic
profiles which produce these isotopes even for initially symmetric
matter. Figure \ref{fig:MC_ye0500_fe60_tc97} shows two such
profiles.

The profile in the top row of Figure \ref{fig:MC_ye0500_fe60_tc97}
has extremum values $T_9^{min}=0.157$, $T_9^{max}=1.01$,
$\rho_{min}=1.177\times10^5$ g cm$^{-3}$, and
$\rho_{max}=1.13\times10^7$ g cm$^{-3}$, and maximizes the
\ux{60}{Fe} yield within our data set. During the non-equilibrium
part of the evolution the QSE cluster dissolves in multiple
small-scale clusters. The dominant yields are determined by
$\alpha$-capture reactions within the mass range $12\lesssim
A\lesssim48$, which is a non-equilibrium and profile dependent
effect. Yields for masses $A\gtrsim60$ are configured by the
interplay of (1) \mr{(n,\gamma)} reactions out of equilibrium which
supply the nuclear flow to $A\gtrsim60$ nuclei at the cost of free
neutrons, (2) \mr{(p,n)} reactions out of equilibrium which
redistribute the nuclear flow among neutron-rich isotopes, and (3)
the small-scale chains of \mr{(p,n)} reactions in equilibrium along
isobars which collect the majority of nuclear flow available. For
instance, equilibrium chains along isobars near \ux{60}{Fe} include
\ux{58}{Cu} to \ux{58}{Fe} for mass $A=58$, \ux{59}{Cu} to
\ux{59}{Co} for mass $A=59$, \ux{60}{Zn} to \ux{60}{Ni} for mass
$A=60$, \ux{61}{Zn} to \ux{61}{Ni} for mass $A=61$, and \ux{62}{Zn}
to \ux{62}{Ni} for mass $A=62$. Although \ux{60}{Fe} does not
participate in the $A=60$ equilibrium chain, it controls the amount
of incoming flow from the $Z=26$ isotopic line that is transferred
to the $A=60$ chain through the reactions
\mr{\ux{60}{Fe}(p,n)\ux{60}{Co}} and
\mr{\ux{60}{Co}(p,n)\ux{60}{Ni}}. Once the free neutrons are
depleted, the mass fractions in this mass range are stabilized until
freeze-out. The top right panel of Figure
\ref{fig:MC_ye0500_fe60_tc97} shows that \ux{79}{Se} is also
produced. Its mass fraction increase close to freeze-out stems from
the action of weak reactions (see Section \ref{sec:reactions}).

The interplay between \mr{(n,\gamma)} and \mr{(p,n)} reactions for
initial $Y_e=0.5$ is another example of non-equilibrium
nucleosynthesis. The formation of small-scale equilibrium patterns
is strongly dependent on the particular combinations of temperature
and density values during the expansion. The profile in the bottom
row of Figure \ref{fig:MC_ye0500_fe60_tc97} looks similar to the
profile in the top row. It has extremum values $T_9^{min}=0.55$,
$T_9^{max}=1.41$, $\rho_{min}=3.4\times10^5$ g cm$^{-3}$, and
$\rho_{max}=4.5\times10^6$ g cm$^{-3}$. The mass fraction curves in
the right column of the Figure have similar trends, but the yields'
distribution is different. \ux{16}{O} is underproduced with respect
to \ux{12}{C} and \ux{28}{Si}, and the production of \ux{97}{Tc} and
\ux{92}{Nb} is favored instead of \ux{60}{Fe} and \ux{79}{Se}. This
is an example of the composition's dependence on the thermodynamic
conditions during the non-equilibrium part of the evolution. It is
noteworthy though that radioactivities in the mass regime
$A\gtrsim60$ are not produced during the exponential and power-law
profiles for $Y_e=0.5$, because the large-scale equilibrium patterns
for most types of freeze-outs (Figure
\ref{fig:contour_AD1_ye0500_a-chain}) favor nuclei primarily within
the silicon and iron groups up to $A\approx56$.

\section{SUMMARY}
\label{sec:summary}

We have used parameterized expansion profiles to explore the details
of nucleosynthesis triggered by two mechanisms during freeze-out
expansions in core-collapse supernovae. The mass layers processed by
each of the two mechanisms are separated by a contact discontinuity
and do not mix. The first mechanism is related to the convection and
instabilities within homogeneous progenitor matter that is accreted
through the supernova shock. The second mechanism is related to the
impact of the reverse shock on the supersonic wind at the inner
regions of the ejecta above the proto-neutron star. The exponential
and power-law monotonic profiles are nucleosynthesis probes for the
cases where the supernova shock (or the reverse shock) raises the
temperature of the progenitor mass layers (or the inner mass layers
respectively) above the NSE threshold and subsequent mixing
processes during the expansion are negligible. Our non-monotonic
profiles aim to simulate thermodynamic trajectories which are
affected by the explosion's asymmetries, instabilities, and mixing
processes following the passage of the supernova shock through the
progenitor mass layers, and the effect of the reverse shock on the
proto-neutron star material when the temperature does not exceed the
NSE or large-scale QSE threshold.

The isotopes produced during the freeze-out expansions are separated
in two families. The first family of isotopes have a region-divided
structure within our parameter space of peak temperatures, peak
densities, and initial electron fraction values. Each region in this
space is associated with a freeze-out type, and its locus depends on
the local equilibrium patterns near the isotope while the large QSE
cluster dissolves. The freeze-out types are characterized by unique
equilibrium state transitions (EST) that the QSE cluster, and hence
the isotopes of the first family, sustain. The mass fraction curves
within a region for the isotopes of the first family are similar,
and their specific profile is shaped by a few critical reactions
which differ from isotope to isotope. These critical reactions are
also responsible for the specific shape of the regions in our
parameter space and their trends from profile to profile, because
they determine whether an EST takes place or not. The first family
includes the majority of isotopes in the mass range $12\leqslant
A\leqslant122$. The isotopes of the second family are related to
nuclei near the magic numbers 28, 50 and 82 and they tend to
dominate the final composition. These isotopes are produced by
maintaining the maximum nuclear flows in their locality and they do
not sustain any EST for all types of freeze-out.

The freeze-out types identified within our parameter space involve
normal, $\alpha$-rich, $\alpha$$p$-rich, $\alpha$$n$-rich, the
chasm, \mr{(p,\gamma)}-leakage and photodisintegration regime.
Freeze-out types are classified according to the EST that the
large-scale QSE cluster sustains, and additional ESTs that isolated
nuclei sustain in local equilibrium patterns. The local equilibrium
patterns are formed once the participating nuclei are left outside
the QSE cluster and are classified in two general categories. The
first category results in the mass fraction configuration of nuclei
until the iron peak. It involves \mr{(p,\gamma)} reaction chains in
equilibrium along isotone lines. If any of the \mr{(p,\gamma)}
reactions along a chain breaks equilibrium, then the isotopes of the
related isotone line sustain their second EST. Dominant yields are
localized near the magic number 28. The second category requires
significant amounts of free neutrons and results in the mass
fraction configuration of nuclei beyond the iron peak until nuclear
masses $A\sim 122$. It involves \mr{(n,\gamma)} reaction chains in
equilibrium, connected with \mr{(p,n)} reactions in equilibrium.
Once the \mr{(p,n)} reactions break equilibrium the isotopes along
the related isotopic lines sustain an EST. Dominant yields are
localized near the neutron magic numbers 28, 50, and 82.

We performed reaction rate sensitivity studies using the exponential
and power-law profiles and utilized non-monotonic expansion profiles
to investigate nucleosynthesis trends of radioactivities from
\ux{26}{Al} to \ux{97}{Tc}. Once produced, these radioactive
isotopes could be transferred to cooler mass layers by mixing
processes during the explosion, and possibly released to the
interstellar medium. Contrary to the exponential and power-law
profiles, non-monotonic expansions involve longer non-equilibrium
nucleosynthesis intervals. The production mechanism details are
strongly dependent on the temperature and density values during the
non-equilibrium part of the evolution, which implies a dependency on
the values of the extremum points for the temperature and density
trajectories. The non-monotonic expansions demonstrate mass fraction
trends and yield distributions that cannot be achieved by the
exponential and power-law profiles. For instance, there are cases
where the silicon group yields are larger than the iron group yields,
which is not possible for monotonic expansion profiles. In addition,
the exponential and power-law profiles tend to produce \ux{60}{Fe},
\ux{79}{Se}, \ux{92}{Nb}, \ux{93}{Zr}, and \ux{97}{Tc} only for
initially neutron-rich composition, while non-monotonic profiles may
produce them even for initially symmetric composition.

\acknowledgments The authors thank Alexandros Taflanidis and
Georgios Varsamopoulos for test calculations and useful discussions.
All calculations have been performed at the High Performance
Computing facilities, operated by the Center for Research Computing
at Notre Dame. G.M. specifically acknowledges the assistance of
Jean-Christophe Ducom, Timothy Stitt, In-Saeng Suh, and Mark
Suhovecky. This work is supported by the NSF under Grant PHY
02-16783 for the Frontier Center ``Joint Institute for Nuclear
Astrophysics'' (JINA).

\clearpage

\bibliographystyle{apj}
\bibliography{traces}

\clearpage

\begin{deluxetable}{lcc}

\tablecaption{553 Isotope Nuclear Network} \tablewidth{100pt}
\tablehead{$Z$ & $A_{min}$ & $A_{max}$} \startdata
H  &  2    &  3   \\
He &  3    &  3   \\
Li &  6    &  7   \\
Be &  7    &  9   \\
B  &  8    &  11  \\
C  &  11   &  14  \\
N  &  12   &  15  \\
O  &  14   &  19  \\
F  &  17   &  21  \\
Ne &  17   &  24  \\
Na &  19   &  27  \\
Mg &  20   &  29  \\
Al &  22   &  31  \\
Si &  23   &  34  \\
P  &  27   &  38  \\
S  &  28   &  42  \\
Cl &  31   &  45  \\
Ar &  32   &  46  \\
K  &  35   &  49  \\
Ca &  36   &  49  \\
Sc &  40   &  51  \\
Ti &  41   &  53  \\
V  &  43   &  55  \\
Cr &  44   &  58  \\
Mn &  46   &  61  \\
Fe &  47   &  63  \\
Co &  50   &  65  \\
Ni &  51   &  67  \\
Cu &  55   &  69  \\
Zn &  57   &  72  \\
Ga &  59   &  75  \\
Ge &  62   &  78  \\
As &  65   &  79  \\
Se &  67   &  83  \\
Br &  68   &  83  \\
Kr &  69   &  87  \\
Rb &  73   &  85  \\
Sr &  74   &  91  \\
Y  &  75   &  94  \\
Zr &  78   &  95  \\
Nb &  82   &  97  \\
Mo &  83   &  98  \\
Tc &  86   &  99  \\
Ru &  89   &  99  \\
Rh &  93   &  99  \\
\enddata
\label{tab:net553}
\end{deluxetable}

\clearpage

\begin{deluxetable}{ccc}

\tablecaption{Isotopes of the Second Family near the Magic Number
28} \tablewidth{100pt} \tablehead{\colhead{$Z$} &
\colhead{$A_{min}$} & \colhead{$A_{max}$}} \startdata
Fe & 56 & 57 \\
Co & 56 & 57 \\
Ni & 56 & 62 \\
Cu & 59 & 63 \\
Zn & 60 & 65 \\
Ga & 63 & 67 \\
Ge & 62 & 69 \\
As & 68 & 71 \\
\enddata
\label{tab:ni56-like_yields}
\end{deluxetable}

\clearpage

\begin{deluxetable}{ccccc}

\tablecaption{Nuclear Reactions Relevant to the Synthesis of
Radioactivities} \tablewidth{480pt}

\tablehead{\colhead{Reaction} &\colhead{Contribution} &
\colhead{$Y_{e}$} & \colhead{Regime} & \colhead{Profile}} \startdata
\multicolumn{5}{c}{$^{26}$Al}\\
\hline
\mr{\ux{26}{Al}_g(p,\gamma)\ux{27}{Si}}  & Chasm formation, depth, shift &  0.50-0.52 & Chasm & Both\\
\mr{\ux{26}{Al}_g(\alpha,p)\ux{29}{Si}}  & Chasm formation, depth, shift &  0.50-0.52 & Chasm & Both\\
\mr{\ux{26}{Al}_g(p,\gamma)\ux{27}{Si}}  & 2nd arc formation/dip &  0.50-0.52 & $\alpha$$p$-rich & Both\\
\mr{\ux{26}{Al}_g(p,\gamma)\ux{27}{Si}}  & 3rd arc formation &  0.52 & $\alpha$$p$-rich & Both\\
\mr{\ux{25}{Al}(e^{-},\nu_{e})\ux{25}{Mg}}  & Flow transfer to symmetric nuclei &  0.50-0.52 & $\alpha$$p$-rich & Both\\
\mr{\ux{25}{Mg}(p,\gamma)\ux{26}{Al}_g}  & 2nd/3rd arc formation &  0.50-0.52 & $\alpha$$p$-rich & Both\\
\mr{\ux{25}{Al}(p,\gamma)\ux{26}{Si}}  & 2nd/3rd arc formation &  0.50-0.52 & $\alpha$$p$-rich & Both\\
\mr{\ux{24}{Mg}(\alpha,\gamma)\ux{28}{Si}}  & Post-arc ascending track &  0.52 & $\alpha$$p$-rich & Power law\\
\mr{\ux{27}{P}(e^{-},\nu_{e})\ux{27}{Si}}  & Flow transfer to symmetric nuclei &  0.50-0.52 & $\alpha$$p$-rich & Exponential\\
\mr{\ux{26}{Si}(e^{-},\nu_{e})\ux{26}{Al}_m}  & Flow transfer to symmetric nuclei &  0.50-0.52 & $\alpha$$p$-rich & Exponential\\
\mr{\ux{26}{Si}(p,\gamma)\ux{27}{P}}  & 2nd arc dip &  0.50-0.52 & $\alpha$$p$-rich & Exponential\\
\mr{\ux{27}{P}(p,\gamma)\ux{28}{S}}  & 3rd arc formation &  0.50-0.52 & $\alpha$$p$-rich & Exponential\\
\mr{\ux{27}{Si}(e^{-},\nu_{e})\ux{27}{Al}}  & Flow transfer to symmetric nuclei &  0.50-0.52 & $\alpha$$p$-rich & Power law\\
\mr{\ux{22}{Mg}(e^{-},\nu_{e})\ux{22}{Na}}  & Flow transfer to symmetric nuclei &  0.50-0.52 & $\alpha$$p$-rich & Power law\\
\hline
\multicolumn{5}{c}{$^{41}$Ca}\\
\hline
\mr{\ux{41}{Ca}(p,\gamma)\ux{42}{Sc}}  & Chasm formation, depth, shift &  0.50-0.52 & Chasm & Both\\
\mr{\ux{41}{Ca}(p,\gamma)\ux{42}{Sc}}  & 3rd arc formation/slope &  0.50-0.52 & $\alpha$-rich  & Both\\
\mr{\ux{40}{Ca}(\alpha,\gamma)\ux{44}{Ti}}  & 2nd/3rd arc dip/slope &  0.50-0.52 & $\alpha$$p$-rich & Both\\
\mr{\ux{41}{Ca}(\alpha,\gamma)\ux{45}{Ti}}  & 2nd arc formation &  0.48 & $\alpha$-rich & Both\\
\mr{\ux{40}{Ca}(p,\gamma)\ux{41}{Sc}}  & 3rd arc formation &  0.50-0.52 & $\alpha$$p$-rich & Both\\
\mr{\ux{41}{Sc}(p,\gamma)\ux{42}{Ti}}  & 3rd arc slope &  0.50-0.52 & $\alpha$$p$-rich & Both\\
\mr{\ux{59}{Cu}(p,\gamma)\ux{60}{Zn}}  & Flow transfer within QSE cluster &  0.50-0.52 & $\alpha$-rich & Both\\
\mr{\ux{37}{Ar}(\alpha,\gamma)\ux{41}{Ca}}  & 2nd arc formation &  0.48 & $\alpha$-rich & Both\\
\mr{\ux{41}{Ca}(\alpha,p)\ux{44}{Sc}}  & 2nd arc formation &  0.48 & $\alpha$-rich  & Both\\
\hline
\multicolumn{5}{c}{$^{53}$Mn}\\
\hline
\mr{\ux{53}{Fe}(e^{-},\nu_{e})\ux{53}{Mn}}  & Flow transfer from symmetric nuclei &  0.50-0.52 & $\alpha$$p$-rich & Both\\
\mr{\ux{52}{Fe}(e^{-},\nu_{e})\ux{52}{Mn}}  & Flow transfer from symmetric nuclei &  0.50-0.52 & $\alpha$$p$-rich & Both\\
\mr{\ux{60}{Zn}(e^{-},\nu_{e})\ux{50}{Cu}}  & Flow transfer from symmetric nuclei &  0.50-0.52 & $\alpha$$p$-rich & Both\\
\mr{\ux{52}{Mn}(p,n)\ux{52}{Fe}}  & Flow transfer from symmetric nuclei &  0.50 & $\alpha$-rich & Both\\
\mr{\ux{42}{Ti}(e^{-},\nu_{e})\ux{42}{Sc}}  & Flow transfer from symmetric nuclei &  0.50-0.52 & $\alpha$$p$-rich & Both\\
\mr{\ux{46}{Cr}(e^{-},\nu_{e})\ux{46}{V}}  & Flow transfer from symmetric nuclei &  0.50-0.52 & $\alpha$$p$-rich & Both\\
\mr{\ux{53}{Fe}(p,\gamma)\ux{54}{Co}}  & Flow transfer from symmetric nuclei &  0.50-0.52 & $\alpha$$p$-rich & Both\\
\mr{\ux{53}{Ni}(e^{-},\nu_{e})\ux{53}{Co}}  & Flow transfer from symmetric nuclei &  0.50-0.52 & $\alpha$$p$-rich & Both\\
\mr{\ux{52}{Fe}(p,\gamma)\ux{53}{Co}}  & Flow transfer from symmetric nuclei &  0.50-0.52 & $\alpha$$p$-rich & Both\\
\hline
\multicolumn{5}{c}{$^{60}$Fe}\\
\hline
\mr{\ux{60}{Fe}(p,n)\ux{60}{Co}}  & Main depletion reaction &  0.48 & $\alpha$$n$-rich & Both\\
\mr{\ux{60}{Fe}(p,n)\ux{60}{Co}}  & 2nd arc formation &  0.48 & $\alpha$$n$-rich & Both\\
\mr{\ux{86}{Kr}(\alpha,n)\ux{89}{Sr}}  & Flow transfer near $N=50$ &  0.48 & $\alpha$$n$-rich & Both\\
\mr{\ux{60}{Fe}(\alpha,n)\ux{63}{Ni}}  & 2nd arc formation &  0.48 & $\alpha$$n$-rich & Both\\
\mr{\ux{59}{Fe}(n,\gamma)\ux{60}{Fe}}  & 2nd arc amplitude &  0.48 & $\alpha$$n$-rich & Both\\
\mr{\ux{58}{Fe}(\alpha,n)\ux{61}{Ni}}  & 2nd arc slope &  0.48 & $\alpha$$n$-rich & Both\\
\hline
\multicolumn{5}{c}{$^{79}$Se}\\
\hline
\mr{\ux{79}{As}(p,n)\ux{79}{Se}}  & 2nd arc formation &  0.48 & $\alpha$$n$-rich & Both\\
\mr{\ux{86}{Kr}(\alpha,n)\ux{89}{Sr}}  & Flow transfer near $N=50$ &  0.48 & $\alpha$$n$-rich & Both\\
\mr{\ux{79}{As}(,e^{-}\;\nu_{e})\ux{79}{Se}}  & Flow transfer at freeze-out &  0.48 & $\alpha$$n$-rich & Both\\
\mr{\ux{78}{Se}(n,\gamma)\ux{79}{Se}}  & Flow transfer before freeze-out &  0.48 & $\alpha$$n$-rich & Power law\\
\mr{\ux{77}{Se}(n,\gamma)\ux{78}{Se}}  & Flow transfer before freeze-out &  0.48 & $\alpha$$n$-rich & Power law\\
\hline
\multicolumn{5}{c}{$^{92}$Nb}\\
\hline
\mr{\ux{92}{Zr}(p,n)\ux{92}{Nb}}  & 2nd arc formation &  0.48 & $\alpha$$n$-rich & Both\\
\mr{\ux{86}{Kr}(\alpha,n)\ux{89}{Sr}}  & Flow transfer near $N=50$ &  0.48 & $\alpha$$n$-rich & Both\\
\hline
\multicolumn{5}{c}{$^{93}$Zr}\\
\hline
\mr{\ux{92}{Zr}(n,\gamma)\ux{93}{Zr}}  & Flow transfer near $N=50$ &  0.48 & $\alpha$$n$-rich & Power law\\
\mr{\ux{86}{Kr}(\alpha,n)\ux{89}{Sr}}  & Flow transfer near $N=50$ &  0.48 & $\alpha$$n$-rich & Both\\
\hline
\multicolumn{5}{c}{$^{97}$Tc}\\
\hline
\mr{\ux{97}{Mo}(p,n)\ux{97}{Tc}}  & 2nd arc formation &  0.48 & $\alpha$$n$-rich & Both\\
\mr{\ux{97}{Tc}(n,\gamma)\ux{98}{Tc}}  & 2nd arc formation &  0.48 & $\alpha$$n$-rich & Both\\
\mr{\ux{98}{Mo}(p,n)\ux{98}{Tc}}  & 2nd arc formation &  0.48 & $\alpha$$n$-rich & Both\\
\mr{\ux{86}{Kr}(\alpha,n)\ux{89}{Sr}}  & Flow transfer near $N=50$ &  0.48 & $\alpha$$n$-rich & Both\\
\enddata
\label{tab:nuclear_reactions}
\end{deluxetable}

\clearpage

\begin{deluxetable}{lccc}

\tablecaption{Comparison between the Monotonic and Non-monotonic
Expansion Profiles} \tablewidth{450pt} \tablehead{\colhead{Features}
& \colhead{Exponential} & \colhead{Power-law} &
\colhead{Non-monotonic}} \startdata
Timescale & Fixed & Dynamic & Dynamic \\
$T^3/\rho$ & Constant & Constant & Non-constant \\
Parameters & 1 & 1 & 6 \\
$T$, $\rho$ coupled & No & Yes & No \\
QSE evolution & Hierarchical & Hierarchical & Periodic \\
Dominant yields & Second family only & Second family only & Both families \\
Radioactivities production & $Y_e<0.5$ & $Y_e<0.5$ & $Y_e\leqslant0.5$ \\
\enddata
\label{tab:profile_comparison}
\end{deluxetable}

\clearpage

\begin{deluxetable}{ccccccc}

\tablecaption{Final Mass Fractions of Radioactivities for
Non-monotonic Expansions} \tablewidth{405pt}
\tablehead{\colhead{$Y_e$} & \colhead{$\rho_0$ (g cm$^{-3}$)} &
\colhead{$T_9^{min}$} & \colhead{$T_9^{max}$} &
\colhead{$\rho_{min}$ (g cm$^{-3}$)} & \colhead{$\rho_{max}$ (g
cm$^{-3}$)} & \colhead{Mass Fraction}} \startdata
\multicolumn{7}{c}{\ux{26}{Al}}\\
\hline
0.48 & $10^7$ & 1.0 & 1.2 & $2\times10^5$ & $1\times10^8$ & $8\times10^{-4}$ \\
0.48 & $10^7$ & 0.04 & 1.54 & $1.4\times10^6$ & $1.2\times10^8$ & $3\times10^{-4}$ \\
0.5 & $10^6$ & 0.12 & 1 & $10^5$ & $1.5\times10^7$ & $1.8\times10^{-3}$ \\
0.5 & $10^8$ & 0.03 & 2.32 & $2.6\times10^5$ & $6.2\times10^8$ & $4\times10^{-3}$ \\
0.5 & $10^8$ & 0.44 & 0.78 & $1.7\times10^5$ & $4.4\times10^8$ & $10^{-3}$ \\
0.52 & $10^7$ & 0.01 & 1.5 & $2\times10^6$ & $10^8$ & 0.2 \\
0.52 & $10^7$ & 0.02 & 0.9 & $2\times10^6$ & $5\times10^7$ & $4\times10^{-3}$ \\
\hline
\multicolumn{7}{c}{\ux{41}{Ca}}\\
\hline
0.48 & $10^7$ & 0.18 & 2.5 & $10^5$ & $5\times10^6$ & $10^{-2}$ \\
0.48 & $10^7$ & 1 & 2 & $2\times10^5$ & $4.3\times10^7$ & $10^{-3}$ \\
0.5 & $10^6$ & 0.04 & 2 & $8\times10^5$ & $4\times10^7$ & $10^{-3}$ \\
0.5 & $10^6$ & 1.7 & 2.3 & $10^5$ & $3\times10^6$ & $10^{-4}$ \\
0.5 & $10^8$ & 1.3 & 2.1 & $1.4\times10^5$ & $6\times10^8$ & $2\times10^{-3}$ \\
0.5 & $10^8$ & 0.98 & 1.95 & $5.7\times10^5$ & $2.4\times10^6$ & $6\times10^{-4}$ \\
0.52 & $10^7$ & 1.8 & 2.4 & $5\times10^5$ & $5\times10^8$ & $10^{-2}$ \\
0.52 & $10^7$ & 0.08 & 0.6 & $2\times10^5$ & $3\times10^6$ & $10^{-3}$ \\
\hline
\multicolumn{7}{c}{\ux{44}{Ti}}\\
\hline
0.48 & $10^7$ & 1.4 & 1.94 & $3.3\times10^5$ & $3.4\times10^6$ & 0.3 \\
0.48 & $10^7$ & 2.3 & 1.78 & $1.9\times10^5$ & $3.8\times10^8$ & 0.2 \\
0.5 & $10^6$ & 0.06 & 2 & $7.1\times10^5$ & $8.1\times10^7$ & 0.4 \\
0.5 & $10^6$ & 0.99 & 1.95 & $1.2\times10^5$ & $1.1\times10^8$ & 0.28 \\
0.5 & $10^8$ & 1.5 & 2.36 & $8.8\times10^5$ & $3.7\times10^6$ & 0.18 \\
0.5 & $10^8$ & 1.77 & 1.88 & $1.5\times10^5$ & $5\times10^8$ & 0.1 \\
0.52 & $10^7$ & 1.7 & 2.3 & $10^5$ & $3\times10^6$ & 0.27 \\
0.52 & $10^7$ & 2.2 & 1.86 & $3.7\times10^5$ & $3\times10^8$ & 0.19 \\
\hline
\multicolumn{7}{c}{\ux{53}{Mn}}\\
\hline
0.48 & $10^7$ & 1.25 & 2 & $10^5$ & $2\times10^6$ & $10^{-2}$ \\
0.48 & $10^7$ & 0.2 & 3.7 & $7.8\times10^6$ & $10^7$ & $10^{-2}$ \\
0.48 & $10^7$ & 0.04 & 3.8 & $1.5\times10^5$ & $3.7\times10^6$ & $10^{-2}$ \\
0.5 & $10^6$ & 0.02 & 2.26 & $1.2\times10^5$ & $4.3\times10^6$ & $4.7\times10^{-4}$ \\
0.5 & $10^6$ & 2.39 & 2.14 & $1.2\times10^5$ & $1.2\times10^7$ & $2.7\times10^{-4}$ \\
0.5 & $10^8$ & 0.07 & 3 & $8\times10^7$ & $2.3\times10^6$ & $4\times10^{-3}$ \\
0.5 & $10^8$ & 1.67 & 2.84 & $4.8\times10^6$ & $1.2\times10^6$ & $1.5\times10^{-3}$ \\
0.52 & $10^7$ & 2.5 & 2.2 & $10^5$ & $2.4\times10^8$ & $6\times10^{-5}$ \\
0.52 & $10^7$ & 0.46 & 1 & $10^6$ & $6.8\times10^8$ & $5\times10^{-5}$ \\
\hline
\multicolumn{7}{c}{\ux{60}{Fe}}\\
\hline
0.48 & $10^7$ & 0.02 & 2.5 & $2\times10^5$ & $3\times10^7$ & $10^{-3}$ \\
0.48 & $10^7$ & 1.5 & 1.3 & $10^5$ & $5.5\times10^6$ & $10^{-3}$ \\
0.48 & $10^7$ & 0.04 & 0.2 & $2\times10^5$ & $2\times10^6$ & $10^{-5}$ \\
0.5 & $10^6$ & 0.04 & 0.54 & $1.7\times10^5$ & $5.8\times10^8$ & $4\times10^{-6}$ \\
0.5 & $10^6$ & 0.04 & 1.14 & $5.5\times10^5$ & $2.4\times10^6$ & $3.7\times10^{-6}$ \\
0.5 & $10^8$ & 0.16 & 1 & $1.2\times10^5$ & $1.1\times10^7$ & $1.9\times10^{-6}$ \\
\hline
\multicolumn{7}{c}{\ux{79}{Se}}\\
\hline
0.48 & $10^7$ & 2 & 1.5 & $10^5$ & $2\times10^8$ & $10^{-2}$ \\
0.48 & $10^7$ & 0.04 & 1.5 & $4\times10^5$ & $2\times10^6$ & $10^{-3}$ \\
0.48 & $10^7$ & 0.04 & 0.03 & $3\times10^5$ & $8\times10^8$ & $6\times10^{-4}$ \\
0.5 & $10^6$ & 0.06 & 1.52 & $5.6\times10^5$ & $2\times10^8$ & $10^{-4}$ \\
0.5 & $10^6$ & 0.06 & 0.6 & $3.8\times10^5$ & $4\times10^8$ & $6\times10^{-5}$ \\
0.5 & $10^8$ & 0.34 & 0.8 & $10^5$ & $7\times10^8$ & $5.6\times10^{-5}$ \\
0.5 & $10^8$ & 0.26 & 2 & $1.2\times10^5$ & $10^8$ & $1.8\times10^{-6}$ \\
\hline
\multicolumn{7}{c}{\ux{92}{Nb}}\\
\hline
0.48 & $10^7$ & 0.1 & 0.5 & $7\times10^6$ & $2\times10^7$ & $2\times10^{-2}$ \\
0.48 & $10^7$ & 0.1 & 1.2 & $6.5\times10^6$ & $4.2\times10^6$ & $2\times10^{-2}$ \\
0.48 & $10^7$ & 0.08 & 2 & $10^6$ & $10^7$ & $10^{-3}$ \\
0.5 & $10^6$ & 0.17 & 1.14 & $1.1\times10^5$ & $9\times10^6$ & $1.9\times10^{-6}$ \\
0.5 & $10^6$ & 0.16 & 0.38 & $2\times10^5$ & $2\times10^7$ & $1.8\times10^{-6}$ \\
0.5 & $10^8$ & 0.17 & 0.29 & $1.7\times10^5$ & $7.4\times10^6$ & $5.8\times10^{-5}$ \\
0.5 & $10^8$ & 0.55 & 1.15 & $10^5$ & $2\times10^6$ & $2.8\times10^{-7}$ \\
\hline
\multicolumn{7}{c}{\ux{93}{Zr}}\\
\hline
0.48 & $10^7$ & 1.7 & 2 & $2.5\times10^5$ & $10^7$ & $4\times10^{-3}$ \\
0.48 & $10^7$ & 0.07 & 1.5 & $2.5\times10^6$ & $1.5\times10^6$ & $3\times10^{-3}$ \\
0.5 & $10^6$ & 0.11 & 0.4 & $5.7\times10^5$ & $2\times10^8$ & $7\times10^{-7}$ \\
0.5 & $10^6$ & 0.11 & 1.1 & $3.5\times10^5$ & $2\times10^7$ & $7\times10^{-7}$ \\
0.5 & $10^8$ & 0.49 & 1.11 & $1.6\times10^5$ & $4\times10^8$ & $4.7\times10^{-7}$ \\
\hline
\multicolumn{7}{c}{\ux{97}{Tc}}\\
\hline
0.48 & $10^7$ & 1.9 & 2.3 & $3.1\times10^5$ & $10^6$ & $5\times10^{-4}$ \\
0.48 & $10^7$ & 2.7 & 2 & $3.8\times10^5$ & $2\times10^7$ & $4\times10^{-4}$ \\
0.48 & $10^7$ & 0.1 & 2.4 & $1.8\times10^6$ & $2\times10^8$ & $2\times10^{-4}$ \\
0.5 & $10^6$ & 0.19 & 1.7 & $1.4\times10^5$ & $1.2\times10^7$ & $7\times10^{-6}$ \\
0.5 & $10^8$ & 0.19 & 0.46 & $1.8\times10^5$ & $5.8\times10^7$ & $1.4\times10^{-5}$ \\
0.5 & $10^8$ & 0.55 & 1.41 & $3.4\times10^5$ & $4.5\times10^6$ & $5.7\times10^{-7}$ \\
\enddata
\label{tab:non-monotonic_yields}
\end{deluxetable}


\clearpage

\clearpage

\clearpage

\clearpage

\clearpage

\clearpage

\clearpage

\clearpage

\clearpage

\clearpage

\clearpage

\end{document}